\documentclass[12pt,preprint]{aastex}

\received{}
\revised{}
\accepted{}
\ccc{}
\cpright{}{}
\newcommand{\bb}{\beta}
\newcommand{\dd}{\delta}
\newcommand{\DD}{\Delta}
\newcommand{\ep}{\epsilon}
\newcommand{\eps}{\varepsilon}
\newcommand{\Ga}{\Gamma}

\newcommand{\ww}{\omega}
\newcommand{\WW}{\Omega}

\newcommand{\Sig}{\Sigma}
\newcommand{\Th}{\Theta}
\newcommand{\lra}{\longrightarrow}
\newcommand{\Ra}{\Rightarrow}
\newcommand{\ra}{\rightarrow}
\newcommand{\unitt}{{\bf  e}_\Th}

\newcommand{\gp}{$\gamma p \lra e^- e^+p$\/}
\newcommand{\gggg}{$\gamma \gamma \lra e^- e^+$\/}
\begin{document}

\title{NEW ENERGY SOURCE CONTROLLED BY GRAVITY ALONE?}

\author{Reva Kay Williams}

\affil{Department of Astronomy, University of Florida, 
P.O. Box 112055, 211 Bryant Space Science Center, Gainesville, 
FL 32611}

\email{revak@astro.ufl.edu}

\begin{abstract}
In this paper, I present results from a theoretical and numerical
(Monte Carlo) {\it N-particle\/} fully relativistic four-dimensional
analysis of Penrose scattering processes (Compton and \gggg) in the
ergosphere of a supermassive Kerr (rotating) black hole.  These general 
relativistic model calculations surprisingly reveal
that the observed high energies and luminosities of quasars and other 
active galactic nuclei, the collimated jets about the polar
axis, and the asymmetrical jets (which can be enhanced by
relativistic Doppler beaming effects) {\it all} are inherent properties
of rotating black holes. From this analysis,  it is shown
that the Penrose scattered escaping relativistic particles
exhibit tightly wound coil-like cone distributions (highly
collimated vortical jet distributions) about the polar
axis, with helical polar angles of escape varying from $0^\circ .5$
to $30^\circ$ for the highest energy particles.  It is  also shown that 
the gravitomagnetic (GM) field, which causes the dragging of inertial 
frames, exerts a force acting on the momentum vectors of the incident 
and scattered particles, causing the particle emission to be asymmetrical 
above and below the equatorial plane, thus appearing to break the  
equatorial reflection symmetry of the Kerr metric. When the accretion disk 
is assumed to be a two-temperature bistable thin disk/ion corona 
(or torus, defining an advection-dominated accretion flow),
energies as high as $\sim 54$~GeV can be attained by these 
Penrose processes alone; and when relativistic beaming is included, 
energies in the TeV range can be achieved, agreeing with observations 
of some BL Lac objects.  When this model is applied specifically to 
quasars 3C~279 and 3C~273, and the Seyfert 1 galaxy MCG---6-30-15, their 
observed high energy luminosity spectra in general can be explained.  
This energy-momentum extraction model can be applied to any size black hole, 
irrespective of the mass, and therefore applies to microquasars as well.  
When applied to the classical galactic black hole source Cygnus X-1,
the results are consistent with observations.  The consistency of these 
Penrose model calculations with observations suggests that the external 
magnetic field of the accretion disk plays a negligible role in the 
extraction of energy momentum from a rotating black hole, inside the 
ergosphere, close to the event horizon where gravitational forces, and 
thus the dynamics of the black hole, appear to be dominant, as would be 
expected.
\end{abstract}

\keywords{acceleration of particles---black hole
physics---galaxies: jets---quasars: 
general---gravitation---relativity}
%\narrowtext
%\mediumtext
%\widetext                                                               
\section{Introduction}
\label{sec:intro}

For almost four decades, since the discovery of quasars, mounting
observational evidence has accumulated that black holes
indeed exist in nature.
Recent observations (Wilms et al.~2001)  of the steep emissivity of 
Seyfert 1 galaxy MCG---6-30-15, indicating strong photon emission
at radii near the event horizon; and observations of the lack of 
evidence of the expected 
ion ``dusty''
torus of M87 (Perlman et al. 2001),
have prompted astrophysicists
to suggest a new energy source.  However, it is hardly a new energy 
source 
to relativists, i.e.,
those who study Einstein's Theory of General Relativity. They knew 
for sometime, at least theoretically, what black holes were 
capable of doing
(Williams 1991, 1995).
Williams (1995, 2003, 2002) shows, through theoretical and 
numerical (Monte Carlo) {\it N-particle\/} 
calculations of Penrose (1969)  
processes, occurring at radii inside the ergosphere of 
a rotating black hole near 
the event horizon: including the ``plunging'' regimes 
(Bardeen, Press, \& Teukolsky 1972; Williams 1995; Krolik 1999), that
the black hole can yield escaping particles with energies up to 
$\sim 54$~GeV.  These particles escape in the form of  
collimated,
symmetrical and asymmetrical jets about the polar axis, confirming
the existence of intrinsically collimated vortical jets, found 
theoretical by de Felice \& Calvani (1972); de Felice \& Curir (1992); 
de Felice 
\& Carlotto (1997); de Felice \& Zanotti (2000): 
from 
geometrical studies of particle
trajectories in a Kerr (1963) metric.  The Kerr metric in general describes 
the spacetime separation of events in the gravitational field of
a rotating compact massive object.

In light of the above observational surprises, particularly  the steep 
emissivity of X-rays producing 
the broad Fe K$\alpha$ emission
line at $\sim 6$~keV in  MCG---6-30-15 (Wilms et al.~2001)
and similar AGNs, it appears 
that gravity has triumphed over proposed forms of electromagnetic 
energy extraction from a black hole, as will be described in this 
paper.
This should be of no surprise near the event horizon, 
where the gravitational forces are so strong that 
electromagnetic radiation itself becomes trapped.

Overall, energy extraction from black holes and the production 
of their associated  
 jets have been the  most poorly understood phenomena
of today.
It is clear that 
 gravitational accretion and 
 magnetic
fields play a role, but how? has been the mystery.  We observe these 
jets 
in quasars and microquasars due
to supermassive and stellar size black holes, respectively.  
Therefore, we know that any effective model must have the commonality
to explain jets in both systems.   
  At present there are two 
popular trains of thought  associated with energy extraction and the
production of jets in black holes:
one is that the jets are inherent properties of geodesic 
trajectories in the Kerr 
metric of a rotating black hole, and thus,
can be described by Einstein's general theory of relativity; 
and the other
is that the accretion disk and its magnetic field through 
magnetohydrodynamics (MHD) are producing the jets.  
Perhaps it could be a combination of the two,  with
gravity  controlling the flow near the event horizon
(Williams 2004),
and MHD controlling the flow at distances farther away.
The observations of the jet of M87 suggest this may be 
the case (Junor, Biretta, \& Livio 1999; Perlman et al.~2001).

There are some proposed MHD model calculations 
using a general relativistic accretion disk that involve 
having the magnetic
field lines of the  disk ``anchor'' to  conductive  
 ionized particles of the disk, inside  the ergospheric region, 
extracting rotational
energy from a Kerr  black hole, by way of a Poynting
flux of electromagnetic energy, out to infinity.  
Such models have been proposed to explain recent observations of 
possible direct evidence for the extraction of energy from a
rotating black hole (Wilms et al. 2001).  
In this paper, however, I point out problems with
such models that make these models highly improbable  to
be at work,  i.e., extracting the energy needed to be consistent with
general observations of sources powered by black holes.
It is agreed by the author that some
form of the Penrose mechanism is employed, but it is argued below
that electromagnetic energy extraction is not an effective 
way to use this
mechanism.  Associated problems with such models are described
in detail in the Appendix.

In a classical paper by Bardeen et al. (1972), 
astrophysical implausible
Penrose processes are discussed concerning the 
 breakup of subrelativistic objects in the ergosphere. However, I
  point out that the ``Penrose-Williams''  
mechanism, described by Williams (1995), involves 
relativistic scattering processes: Such processes can be 
very efficient (Piran, Shaham, \& Katz 1975; Williams 1995), 
and do not fall under the 
category of being implausible due to hydrodynamical constraints
(Bardeen et al. 1972),  
since the incident and target particles in the collisions are 
already relativistic, having speeds
$\sim c$. 

 The Penrose mechanism as described here 
(Williams 1995) has a 
``one-on-one'' consistent 
relationship with accretion disk particles.  For example, 
particles from the accretion 
disk can populate the high energy gravitationally blueshifted 
trapped orbits
(or so-called plunging orbits)
at $r<r_{\rm ms}$, the marginal stable orbit (Bardeen et al. 1972).   
Particles
in these now populated orbits can undergo Penrose processes with 
lower soft X-ray energy 
infalling
accretion disk photons:  Penrose Compton scattering (PCS) produces 
copious distributions of high energy X-rays and soft $\gamma$-rays, and 
Penrose pair production (PPP) (\gggg)  produces
copious distributions of relativistic
electron-positron  ($e^-e^+$) pairs, with up to $\sim 90$\% of the 
 particles escaping along vortical 
orbits (\S~\ref{sec:collimation}) that circle the polar axis of the 
KBH many times, as spacetime itself is  dragged around because of
gravity (Williams 2004).  The particles
escape to infinity along well defined four-momentum 
trajectories, with some  intersecting 
the disk (i.e., returning to be reprocessed and/or escaping
to infinity).  This scenario is particularly consistent with  recent 
observations 
of MCG---6-30-15 (\S~\ref{sec:MCG}), and other black hole sources
(\S~\ref{sec:observation}).

Importantly, in these Penrose processes we do not need the  
magnetic field of the accretion disk to ``communicate'' between the 
accretion disk and the 
black hole.  Therefore, there is no need for the 
Blandford and Znajek (BZ) (1977) proposed type models (and
their many associated problems) in the direct  
role of energy extraction from a spinning black hole.  
However, their presence appears to be need once particles are on
escaping orbits, serving the same effects they do in the jets of 
protostars, i.e., appearing to have a dominant role on a large scale,
within the weak field limit,
at distances outside the strong effects of general relativity.
 
As for producing the observed
synchrotron radiation, indicating the present of a magnetic field
near the core region, this radiation 
could very well be produced by the intrinsically self-induced 
magnetic field due to the 
dynamo-like
action of the escaping Penrose produced $e^-e^+$ pairs, escaping
on vortical,
coil-like  trajectories concentric the polar axis, in the form of a 
swirling ``current'' plasma.  This, therefore, adds 
more to the 
unimportance  of the accretion disk magnetic field near the 
event horizon.   

Moreover, although  suggested to be  evidence of 
rotational magnetic energy 
extraction from the Seyfert 1 galaxy MCG---6-30-15 (Wilms et al. 2001),
it appears, as we shall see in this 
paper,  that it is 
gravitational energy momentum being extracted, in the form of 
a relativistic particle flux 
via Penrose processes, as
 described by 
Williams (1995),  
and {\it not} the Poynting flux of electromagnetic 
energy suggested: produced by magnetic field lines 
torquing the black hole or 
plunging accretion disk material,
as described by the BZ-type models.
In the Penrose-Williams mechanism, the steep emissivity profile
[$\varepsilon (r)\sim r^{-5}$] of X-ray photons observed 
(Wilms et al. 2001), requiring a X-ray source that is both powerful 
and very centrally concentrated (which cannot be explained by
standard accretion disk models),
is consistent with energy being extracted by Penrose Compton
scattering processes,
occurring at radii between the marginally bound and marginally stable
orbits, $r_{\rm mb}$ and $r_{\rm ms}$, respectively (Williams 1995).
This black hole source MCG---6-30-15 will be discussed further 
in \S~\ref{sec:MCG}.

Nevertheless, 
once these Penrose processes have occurred and particles are on escaping 
trajectories, they can then interact, say with the expected 
large scale structure disk
magnetic field, at some effective radius $r$ where this
field becomes important
in jet collimation, probably similar to a relative radius 
existing for  the
collimated bi-polar jets of
protostars, which, as mentioned above,
appear to be  undergoing some type of BZ effect---the
direct effect is still somewhat unclear.
  It appears that the magnetic 
field of the accretion disk serves to aid in 
collimating into jets  gravitational binding energy release due to
gravitational accretion, in both protostars and
AGNs (or microquasars); however, in the latter it appears that 
the jets are superimposed
with collimated particles from Penrose processes.

So, overall, in this paper, an analysis of the Penrose mechanism 
is presented to 
describe gravitational-particle interactions close to the event horizon
 at radii $<r_{\rm ms}\simeq 1.2M$ 
and down to the {\it photon orbit},
$r_{\rm ph}\simeq 1.074M$, for a canonical KBH with $a=0.998M$ 
(Thorne 1974), where $a$ is the angular momentum  per unit mass parameter.
In this fully general 
relativistic description, polar jets of 
relativistic particles
of photons and $e^-e^+$ pairs are
produced and collimated by gravity alone, without
the necessity of the  external magnetic field of the accretion
disk.   This theoretical
and numerical model of Penrose processes
 can apply to any size black
hole, and suggests a complete theory for the extraction of 
energy momentum from 
a rotating black hole.  In \S~\ref{sec:model} a summary of the general 
formalism of the model is presented.  In \S~\ref{sec:results},
results of theoretical and
numerical calculated luminosities and energies are presented, 
along with discussion of the escaping
particles'  space momentum trajectories: featuring asymmetrical polar 
distributions and vortical orbits.  Also in \S~\ref{sec:results}, 
agreement with 
observations of specific sources are presented.  Finally, in 
\S~\ref{sec:conclusions} a summary and conclusions are presented.

\section{Model Formalism}
\label{sec:model}

The primary model (Williams 1995) consists of a 
supermassive $10^8 M_\odot$ rotating Kerr
 black hole
plus particles from an
assumed
relativistic bistable thin disk/ion corona [or torus, i.e., 
advection-dominated accretion flow (ADAF)],
two-temperature [separate
temperatures for protons ($\sim 10^{12}$~K) and
electrons ($\sim 10^9$~K)] 
 accretion (Williams 2003; Novikov \& Thorne 1973; 
Eardley \& Lightman 1975; Eilek 1980; Eilek \& Kafatos 1983).   
The bistable accretion disk can exist either in the thin disk
phase and/or the ion corona (ADAF) phase, or oscillate between the 
two (see  Williams 2003 and references therein) 
in various degrees---which could be 
responsible for the observed
variability.  
The Penrose effect as employed
here can operated in either phase.
 The Penrose mechanism
is used to extract rotational energy momentum by
scattering processes inside the ergosphere
($r_{\rm 0}\simeq 2M$, in the equatorial plane for $a=0.998M$).
See Williams (1995) for a detailed description of the
model. 
The
``quasi-Penrose'' (Williams 1991, 1995) processes investigated are
(a) Penrose Compton scattering (PCS) of equatorial
low energy
radially infalling
photons by equatorially confined ($Q_e=0$) and
nonequatorially confined ($Q_e\neq 0$) orbiting target electrons, 
at radii between
the marginally bound ($r_{\rm mb}\simeq 1.089M$) and marginally stable
($r_{\rm ms}\simeq 1.2M$) orbits, where $Q_e$ is the so-called Carter 
constant (Carter 1968), referred to as the $Q$ value (Williams 1995);
(b) PPP (\gp) at $r_{\rm mb}$; and
(c) PPP (\gggg) by equatorial low energy
radially infalling
photons and high energy gravitationally blueshifted 
(by factor $e^{-\nu}\simeq 52.3$)
nonequatorially
confined $\gamma$-rays
at the photon orbit ($r_{\rm ph}\simeq 1.074M$), where 
$e^{-\nu}$ is the ``blueshift'' factor given by the $g_{tt}$ 
component of the 
Kerr metric (see Williams 1995).  Although in the scattering plane
the incident angle of the infalling photon relative to the target
particle is expected in general
to be at least between $0^\circ-90^\circ$ (due to the bending of 
light and/or inertial frame dragging),
maximum energy is extracted in the process when the incident
angle is $90^\circ$, as it is for radially infalling photons 
($L=P_\Phi=0$, where $L$ is the azimuthal coordinate angular momentum).
 Note, the target particles are initially in bound (marginally stable
or unstable) trapped orbits, trapped in the sense of possibly having
no other way of escaping save these Penrose processes 
(Bardeen et al. 1972; Williams 1995).  
Note also that, as the nonequatorially
confined target particle, whose orbital trajectory is derived 
by Williams (1991, 1995; see also Williams 2002), passes 
through the equatorial plane, in its
bound circular orbit at constant radius,
the $Q$ value, a constant of motion as measured by 
an observer at infinity
(Carter 1968; Williams 1995), equals $P_\Th^2$,
where $P_\Th$ is the polar coordinate momentum of the particle.
Setting $P_\Theta=0$, in the Carter constant expression for the
orbital $Q$ value, gives the maximum and minimum latitudinal angles
of the trajectories about the equatorial plane for Wilkins' (1972)
``spherical-like'' nonequatorially confined orbits (see Williams 2003).
These unstable, bound or marginally bound orbits (equatorial,
nonequatorial) of the target particles are assumed to be populated by
accretion disk instability processes and prior Penrose processes.
Such particles must satisfy conditions
 to have a turning point
at the scattering radius (note, a bound stable orbit is considered to
have a ``perpetual'' turning point).  
These conditions depend on the
orbital conserved parameters of the particle: $E$, the energy, and $L$,
or $Q$.  In Williams
(1995, 2002) such conditions are discussed in detail; see
also the possible scenario discussion in \S~\ref{sec:asymmetry} of 
this present manuscript.   In addition, the ``instability phase,''
during which the target particle orbits are presumed to be populated, 
could very well be related to the timescale of the prominent observed 
variabilities of the source. 

Radial infalling equatorially confined incident photons are assumed,
not only for maximum energy extraction but for the simplicity of
their geodesics as well,
since it appears that an infalling equatorially confined photon will not
acquire gravitationally blueshifted orbital energy momentum
 as measured by an observer at infinity, only frame dragging 
blueshifted energy (eq.~[2.8$d$] of Williams 1995).  This is
because the $Q$ value of such photons is zero (see eq.~[2.27] of 
Williams 1995).
The incoming photons, however, need not be confined to the equatorial plane.
In these calculations if equatorially confined infalling photons were
not desired, $\sqrt{Q_{\rm ph}}\equiv (P_{\rm ph})_\Theta$ of the 
initial photon would not be set equal zero.  That is, the model 
calculation is set up such that one can change the initial 
energy-momentum four vector components (or four-momenta) of the 
incident and target particles to accommodate any 3-space dimensional 
geometrical disk configuration.
Moreover, in an ADAF (including the ion corona), during the infall of 
particles, through the ergosphere, some of the particles are expected
to become trapped in nonequatorial ``spherical-like orbits'' (Wilkins 
1972): such orbits would past through the equatorial plane: here is 
where the scattering takes place in these calculations.
Note, the target photons  at the photon orbit can only exist
in nonequatorially confined orbits (Williams  1995); this is also pointed
out by Bardeen (1973).

Monte Carlo {\it N-particle\/} computer simulations of up to
$\sim 70,000$
scattering events of
infalling accretion disk photons (normalized to a
power-law distribution)  are executed for each computed 
Penrose produced luminosity
spectrum (Williams 2003).
Energy and momentum (i.e., four-momentum)
spectra of escaping particles
($\gamma$-rays, $e^-e^+$ pairs), as measured by an observer at infinity,
are obtained per each 2000 scattering events per monochromatic infalling
photon distribution.
The following constituents are used (Williams  1995):
(1) General relativity is used [the Kerr metric spacetime
geometry yields  equatorially and nonequatorially confined
spherical-like (Wilkins 1972) particle
orbits and escape conditions, conserved energy and angular momentum
parameters, and transformations from the Boyer-Lindquist
coordinate frame (BLF) to the local nonrotating frame (LNRF)].
Note, BLF is the observer at infinity (Boyer \& Lindquist 1967);
LNRF is the local Minkowski (flat) spacetime.
(2) Special relativity is used [in the LNRF, physical
processes (i.e., the scatterings) are done; Lorentz transformations
between inertial frames are performed;
and Lorentz invariant laws are applied].
(3) Cross sections are used [application of the Monte Carlo method 
to the
cross sections, in the electron rest frame for PCS, in the proton
rest frame  for PPP(\gp), and in the center of momentum frame
for PPP(\gggg), give the  distributions of scattering angles and final
energies].

\section{Overall Results and Discussion}
\label{sec:results}

\subsection{Energy and Luminosity Spectra Extracted}
\label{sec:energy-luminosity}

In general, energies attained using the proposed
accretion disk model are the following (Williams  1995):

I. {\it \/PCS}.---For the input photon energy 
range $\sim 0.511$~keV to  0.15~MeV, the corresponding
output energy range is $\sim$ 3~keV to 7~MeV. 
The input photon range covers the range of photons in a
thin disk ($\sim 0.511-3.5$~keV), thin
disk/ion corona ($\sim 0.511-3.5$~keV, $\sim 30-150$~keV),
and ADAF ($\sim 30-150$~keV) for a $10^8 M_\odot$ KBH (Williams 2003,
1995).  The input luminosity spectra are based on observations,
consistent with a power-law distribution in the X-ray, and accretion
disk theory.
Typical output luminosity  spectral distributions from PCS are
displayed in Figure 1$a$, the curves passing through numbers $1-13$
(as will be described below in the discussion of the model produced 
luminosity).

II. {\it \/PPP} (\gp).---There are no escaping pairs for 
radially infalling equatorially confined
$\gamma$-rays ($\sim $40~MeV) and no energy boost: implying that
the assumption: negligible recoil energy  given to the proton,
made in the conventional cross section, and perhaps  
the geometry of the
scattering must
be modified.
It had been predicted (Leiter \& Kafatos 1978)
 that pairs with
energies ($\sim $1 GeV) could escape.  See Williams (1995) for
further details of this PPP process.

III. {\it \/PPP} (\gggg).---An input photon energy 
range $\sim$ 3.5~keV to 200~MeV yields
output ($e^-e^+$) energy range $\sim$ 1~MeV to 10~GeV
(for a proton Maxwell-Boltzmann distribution),
and higher up to
$\sim$ 54~GeV (for a proton power-law  distribution, 
with input photon energy $\sim 2$~GeV), where
Maxwell-Boltzman and power-law distributions are
for the  accretion disk protons: undergoing nuclear proton-proton
scatterings,  which yield  neutral pion decays
$\pi^0\lra\gamma\gamma$ (Eilek 1980; Eilek \& Kafatos 1983;
Mahadevan, Narayan, \&  Krolik 1997) that can possibly
populate the photon orbit.  Below, I refer to such decay produced
photons  and 
subsequent $e^-e^+$ pair production (from such 
photons), which can occur in
ADAFs, as Eilek's particles (Eilek 1980).

Specific disk model correlations are the following 
[see Williams  (1995, 2003)
for further details]:

\noindent 
1. Without instabilities
[implying the classical thin disk (Novikov \& Thorne 1973)]:
\begin{itemize}
\item[ $a$)]PCS can convert infalling (incident) soft X-rays 
0.511$-$3.5~keV to moderate X-rays, escaping with energies in 
the range  $\sim$~3$-$262~keV.
The upper and lower bounds on the energy of the outgoing photons are set
by the initial four-momentum conditions of the target orbiting 
electron (with $E_e\simeq 0.539$~MeV at $r_{\rm mb}$) and the 
incident photon
(with $E_{\rm ph}=0.511-3.5$~keV) undergoing  PCS.  These 
initial four-momenta are consistent in general with the following:
theoretical accretion disk models, the threshold energy values for the
scattering process to occur, and what brings about the most ``efficient''
energy extraction process [see Williams (1995) for details defining the
efficiencies].  These initial momenta are substituted into appropriate
theoretical analytically derived model equations 
(describing the Penrose scattering process in the ergosphere of a 
Kerr black hole; see Williams 1995), and the equations are computed.  
The output energy 
range presented above gives the lowest and
highest energy values obtained by the escaping Penrose Compton
scattered (PCS) photons, 
for the given input energy range of the incident photons.
\item[$b$)]Inwardly  directed PCS photons that have an
appropriate turning point
(see Williams 2002)
can serve as seed $\gamma$-rays for the PPP (\gggg)  at the
photon orbit. Specifically, the PCS photons that satisfy  
conditions to have a turning point, acquire gravitationally 
blueshifted energies as high as $\sim 7$~MeV.
\item[$c$) ]PPP (\gggg) can convert infalling soft X-ray photons to
relativistic $e^-e^+$ pairs, escaping with energies in the range
$\sim$ 2$-$6~MeV, i.e., infalling photons can pair produce at the 
photon orbit with photons populated by prior PCS.
\end{itemize}

\noindent
2. With instabilities (implying the
thin disk/ion corona or ADAF):
\begin{itemize}
\item[$a$)] PCS can convert infalling X-rays 0.03$-$0.15~MeV
to escaping photon energies in the
range $\sim$ 0.4$-$7~MeV.  The differences in the 
calculations of the above case in item~1 and
the present case of item~2 are the energies of the
infalling incident photons and the target electrons
($E_e\simeq 0.539-4.8$~MeV), including nonequatorially confined
targets, with $0.963~{\rm MeV}\la E_e \la 4.8$~MeV, for
$\pm 1.68Mm_e\leq (P_e)_\Theta\equiv\sqrt{Q_e}\leq\pm
10.05Mm_e$, corresponding to 
ADAFs with $30~{\rm keV}\la kT_e\la 150~{\rm keV}$, respectively,
after being gravitationally blueshifted at $r_{\rm mb}$ by factor
$e^{-\nu}\simeq 32$,
where $(P_e)_\Theta$ is the polar coordinate momentum component
as measured by an observer at infinity (i.e., in the BLF; see 
\S~\ref{sec:model}).  [Note,
the conserved energy $E=E(Q)$ and azimuthal angular momentum 
$L=L(Q)$ of the target nonequatorially confined test particle orbits
are given by analytically derived expressions presented in 
Williams (2004, 2002, 1995).]  The accretion
disk model, used, is discussed in detail in Williams (2003; see Figure~1
and Table~1 of that reference).  
In general, the target electron orbits are
assumed to be populated during instability phases, more or less, in both
the thin disk and thin disk/ion corona (or ADAF).  In 
\S~\ref{sec:asymmetry}  in the
discussion of a possible scenario for ``jet reversal,'' a brief
description is included on populating the target electron orbits from the
inner region of a thin disk (Novikov \& Thorne 1973). 
\item[$b$)]PCS photons that can populate the photon orbit
(as in item 1.$b$)
have gravitationally blueshifted energies specifically in the range 
$24-57~{\rm MeV}\la E_{\rm ph}^{\prime\prime}\la 348$~MeV,
where $E_{\rm ph}^{\prime\prime}$ is the energy at the photon
orbit due to prior PCS (Williams 2002).  The lower limits of 
$E_{\rm ph}^{\prime\prime}$ are due to PCS by equatorially
[$E_e\simeq 0.539$~MeV, $(P_e)_\Theta=0$] and nonequatorially
[$E_e\simeq 0.96$~MeV, $(P_e)_\Theta=\pm 1.68Mm_e$] confined electron
targets, respectively: note, as in item 2.$a$, these nonequatorially 
confined
target electrons are assumed to come from an ADAF ($kT_e\sim 30$~keV).
\item[$c$)]PPP (\gggg) can convert infalling
soft X-ray photons to relativistic $e^-e^+$ pairs,
escaping with energies in the range $23-56~{\rm MeV}\la E_\mp
\la 340$~MeV. 
Note, the ``stability'' of a turning point being 
perpetual (i.e., bound) at the photon orbit decreases 
with increasing energy of the incoming
incident photons undergoing PCS by equatorially 
confined target electrons.  These calculations show that 
the most stable orbits (or turning points) appear to be 
the ones in which the infalling incident photons and the orbiting 
target electrons are self-consistent, i.e., of 
the same accretion disk phase (e.g., thin disk or ADAF).
\end{itemize} 

\noindent
Note, whenever the thin disk is present, the processes described 
above in items 2.$a$$-$2.$c$ will occur in addition to those 
described in items~1.$a$$-$1.$c$.

\noindent
3. With instabilities [implying the
thin disk/ion corona model or ADAF plus Eilek's  
particles (Eilek 1980; Eilek \& Kafatos 1983) 
to populate the target particle orbits, of electrons
and photons, particularly
the large $Q$-value orbits; see Williams  (1995)]:
\begin{itemize}
\item[$a$)]PCS  can convert infalling photons
0.03$-$0.15~MeV to escaping energies in
the range $\sim$ 6$-$14~MeV.   This is in addition to the 
energy distribution of escaping PCS photons given in item 2.$a$.  
Eilek's particles contribute to the ion 
corona, nonequatorially confined
$e^-e^+$ pairs with energies peak around $E_e\sim 35$~MeV.
At the peak, such electrons with inward trajectories ($P_r<0$)
would have to satisfy  conditions to have a turning point at 
$r_{\rm mb}\la r\la r_{\rm ms}$ (Williams 2004),
requiring, say for the scattering radius
$r_{\rm mb}$,  $\sqrt{Q_e} \ga 73 Mm_e$ and/or 
$L_e\ga 141Mm_e$; these electrons do not appear to be important
in the PCS process.  Observations suggest, however, that 
PCS by Eilek's nonequatorially
confined $e^-e^+$ pairs with $E_e\sim 6-12$~MeV, yielding escaping
energies in the range given above,  might be important, 
requiring, for turning
points to exist at $r_{\rm mb}$,  $\sqrt{Q_e} \ga 
12-25 Mm_e$ and/or $L_e\ga 24-48 Mm_e$, respectively 
(compare Fig.~1$a$; see also
Williams 2003).  
[Note, Eilek's
electrons $\sim 35$~MeV may be an important source of synchrotron 
emission into the 
IR, for a magnetic field strength $B\sim 10^2$~G
(see \S~\ref{sec:asymmetry}).] 
\item[$b$)]PPP (\gggg) can convert infalling soft 
X-rays to relativistic
$e^-e^+$ pairs, escaping with energies ranging from $\sim$ 300~MeV to
as high $\sim 10$ GeV~[for a proton Maxwell-Boltzman distribution
 (see item~III above)], with
input photon energy $\sim 6-200$~MeV from $\pi^0$ decays
(Eilek \& Kafatos 1983).   That is, the input (target) photons
are gravitationally blueshifted (by factor $e^{-\nu}\sim 52$) at 
the photons orbit 
to energies
$E_{\gamma 2}\sim 312~{\rm MeV}-11~{\rm GeV}$, and are assumed to have 
a turning point at (or near) this scattering radius, with 
$\sqrt{Q_{\gamma 2}}\sim 9-312 Mm_e$, respectively, where the subscript 
$\gamma 2$ 
represents the orbiting target photon.
These PPP (\gggg) processes occur in additions
to those given in item 2.$c$. The exact range of the PPP electrons 
will depend on
which of the inwardly directed photons, after being 
gravitationally blueshifted,  satisfy  conditions
to have a turning point at or near the photon orbit [see
Williams (2002) for details].  For completeness, if protons of
Eilek's particles have a power-law distribution 
(Mahadevan et al. 1997) as mentioned in item III, the maximum 
energy attained, using the 
Penrose-Williams' model, for the 
escaping PPP (\gggg) electrons, is $\sim 54$~GeV, for input 
photon energy $\sim 2$~GeV, after being gravitationally blueshifted to 
$E_{\gamma 2}\sim 108$~GeV, with 
$\sqrt{Q_{\gamma 2}}\sim 3121 Mm_e$.
\end{itemize}

Note that, there will be a slight time delay between PCS and 
PPP (\gggg) in
items 1.$a$$-$1.$c$ and 2.$a$$-$2.$c$ that might be consistent with
 the time 
 offset ($\sim 5$~min) between
X-ray and IR flares observed in microquasar GRS 1915+105, indicating 
that these flares are produced by the same event: The X-ray 
flares occur with the 
apparent disappearance of the inner X-ray emitting 
region of the accretion disk; and the 
subsequent IR flares are proposed to be due to synchrotron emitting
ejecta of relativistic plasma into the polar direction 
(Eikenberry et al. 1999a; Eikenberry et al. 1999b).

Before discussing the luminosity spectra produced by these Penrose
processes, we first discuss the ``characteristic voids,'' 
existing, in general, in observed spectra of all AGNs, 
more or less
(compare Fig.~1$a$), and how these Penrose processes suggest an 
explanation for them.   These observed voids appear to be caused
by the ``transitional energy regime'' between thin disk 
($E\sim 0.511-3.5$~keV) and ion corona (or ADAF) ($E\sim 30-150$~keV)
states (see Table~1 of Williams 2003):
therefore, we expect the Penrose process to be void of participating
particles with energy in the range $3.5~{\rm keV}<E<30$~keV, i.e.,
if we assume such particles are sufficiently short-lived,
save the infalling disk electrons with $E\sim 17$~keV that can 
satisfy conditions to populate
the equatorially confined target electron orbits ($Q_e=0$) at radii 
$r\sim r_{\rm mb}$ (see \S~\ref{sec:asymmetry} and paragraph below),
where $E \equiv E_e\sim E_{\rm ph} $ indicates a 
general particle energy, predicted theoretically by a 
particular phase of the accretion disk.  
The PCS photon energies $E_{\rm ph}^\prime$ produced by incident  
and target particles in the transitional energy regime, and the 
subsequent gravitationally blueshifted energy 
$E_{\rm ph}^{\prime\prime}$ at the photon orbit, 
of incoming PCS
photons satisfying  conditions to have a turning point there
(Williams 2002), expected to undergo PPP (\gggg), are found to 
give characteristic voids in the following regimes:
\begin{itemize}
\item[($a$)]For thin disk/ion corona ($kT_e\sim 30$~keV):
$0.262~{\rm MeV}<(E_{\rm ph}^\prime)_{\rm void}<1.1$~MeV 
and $7.3~{\rm MeV}<(E_{\rm ph}^{\prime\prime})_{\rm void}
<57$~MeV, where
the upper limit originates from nonequatorially
confined target electrons, consistent with the electron 
temperature in the ion corona (see Williams 2003).  
Compare with item 2.$b$ above. 
\item[($b$)]For thin disk/ion corona ($kT_e\sim 50$~keV):
$0.262~{\rm MeV}<(E_{\rm ph}^\prime)_{\rm void}<2.1$~MeV
and $7.3~{\rm MeV}<(E_{\rm ph}^{\prime\prime})_{\rm void}<107$~MeV. 
\item[($c$)]For thin disk/ion corona ($kT_e\sim 150$~keV):
$0.262~{\rm MeV}<(E_{\rm ph}^\prime)_{\rm void}<5.3$~MeV
and $7.3~{\rm MeV}<(E_{\rm ph}^{\prime\prime})_{\rm void}<273$~MeV.                  
\end{itemize}

Note, the PCS processes
considered above are those occurring at or near $r_{\rm mb}$:
since the highest energy will be extracted from this scattering 
radius, and it appears that the orbits at this radius will be the
first to be populated, as the disk temperature increases
(\S~\ref{sec:asymmetry}), i.e.,  
because of the larger energy blueshift factor acquired, and the 
smaller $Q_e$ needed,
relative to these parameters at $r_{\rm ms}$.
The disk electron energies $\ga 30$~keV but $< \mu_e$,  being
 gravitationally blueshifted by a factor $e^{-\nu}\sim 32$
(see \S~\ref{sec:model}), satisfying
appropriate turning point conditions (Williams 2004), 
with $Q_e>0$, are assumed to 
populate the  nonequatorially confined target particle orbits for PCS 
(see also \S~\ref{sec:asymmetry}), where $\mu_e\simeq 0.511$~MeV is
the rest mass energy of an electron.  A relativistic four-momentum 
treatment of disk particle processes in thin disk/ion corona 
accretion, inside the ergosphere, appears to be  needed to theoretically
validate this plausible assumption: at  present, however, we do not have 
such a model; therefore, we must rely on what observations convey to us.
 
Moreover, PCS by equatorially confined electron targets ($Q_e=0$), 
assuming to originate
from ``mild'' instabilities in the thin disk (that would cause 
the electron energy to increase to $\sim 17$~keV, however, 
while still predominantly in the thin disk phase) and
radially infalling photons confined along the equatorial plane, 
originating
from the ion corona ($E_{\rm ph}\ga 30$~keV), are not included in the 
above consideration
of the characteristic  voids.  The reason for the exclusion is
that observations suggest such PCS may not be important (compare
Fig.~1$a$,  curve between points 6 and~7), which could mean that 
these target orbits are depopulated while the disk is in the thin disk
or transient phase, and therefore not available for PCS in the ion corona
(ADAF) phase.
Further, for the subsequent gravitationally
blueshift of such inward directed PCS photons, with $E_{\rm ph}^\prime\sim 
0.5$~MeV, $\sim 0.7$~MeV, $\sim 1.6$~MeV,   
corresponding to potential turning  point energies:
$E_{\rm ph}^{\prime\prime}\sim 24$~MeV, $\sim 34$~MeV, $\sim 51$~MeV,
at the photon orbit, for the ion
coronas in items ($a$)$-$($c$), respectively,  we find that most
of these potential turning points, it seems, are
``highly'' uncertain (based on whether 
$Q_{\rm ph}^\prime\simeq Q_{\gamma 2}$ for 
$E_{\rm ph}^{\prime\prime}=e^{-\nu}E_{\rm ph}^\prime\equiv E_{\gamma 2}$). 
Nevertheless, as the energy of the trapped target electron 
is
increased ($Q_e\neq 0$), consistent with the general ion corona electron 
temperature,
the uncertainty of the turning point orbit, being true, decreases;
then $(E_{\rm ph}^\prime)_{\rm void}$ and 
$(E_{\rm ph}^{\prime\prime})_{\rm void}$ are as given above in items 
($a$)$-$($c$).  

Finally, in the above characteristic voids, PCS involving
thin disks with energies less than 
$E\sim 3.5$~keV, and PCS involving Eilek's nonequatorially 
confined $e^-e^+$ pairs,
possibly occurring in the ion corona or ADAF (particularly $\sim 
6-12$~MeV, the range, based on observations, that appears to be the regime 
satisfying turning point conditions), are not included.  
Inclusion of these would slightly  
affect the voids, yet the distinctive characteristics 
would remain.  We will return to this discussion of the characteristic 
voids later in this section.                                                

The luminosity spectrum due to Penrose
processes for the specific case of quasar 3C~273 is plotted in
Figure~1$a$, along with the observed spectrum
for comparison (heavy solid curves superimposed with squares or 
dots).
The outgoing (escaping) luminosity spectrum produced by the
Penrose scattered particles
is given by (Williams 2003)
\begin{eqnarray}
L_\nu^{\rm esc}&\approx&4\pi d^2 F_\nu^{\rm esc}~~
({\rm erg/s\,Hz}) \nonumber\\
                &\approx&4\pi d^2h\nu^{\rm esc}
                f_1 f_2 \cdots f_n\,(N_\nu^{\rm in}-N_\nu^{\rm cap}),
\label{eq:lum}
\end{eqnarray}
where $d$ is the cosmological distance
of the black hole source;
$F_\nu^{\rm esc}$ is the flux of escaping photons;
$N_\nu^{\rm in}$ and $N_\nu^{\rm cap}$ are the emittance of
incoming and
captured photons, respectively;
the $f_n$ values
define the total fraction of the particles that undergoes scattering
[$n=2$ for PCS and $n=5$ for PPP (\gggg)].
The values of $f_1,\ldots, f_n$ are the fitting factors,
which can make the Penrose calculated
luminosities agree with observations
for the specific case of 3C~273, to account for in general our letting 
every particle scatter in the model calculations, since in a realistic 
situation every 
particle will not scatter.
In short, the $f_n$ values, defined as somewhat free parameters,
are probabilities, which are $\leq 1$, but $> 0$;  they are
dependent on the cross sections---for PCS and PPP 
($\equiv f_2$, $f_4$), the fraction of
the luminosity from the disk intersecting the scattering radii
($\equiv f_1$, $f_3$),
and the expansion rate of the jet ($\equiv f_5$).
Note, from equation~(\ref{eq:lum}) we obtain the model calculated 
continuum emission
given by the top curves on Figure~1$a$ (labeled
with numbers for specific cases of target and incident particles; see 
below) 
if we  allow $f_1=f_3\sim 10^{-2}$,  and set the remaining
$f_n$'s  equal 1, where we are assuming that the polar angle 
subtending the bandwidth, $\Delta\theta$,
straddling the equatorial plane, impinged by the luminosity, is 
$\sim 2^\circ$ at $r_{\rm mb}$ and
$\sim 1^\circ$ at $r_{\rm ph}$). 
See Williams (2003) for further details and complete 
definitions of
the $f_n$ values.   The spectrum resulting from
the PPP (\gggg) is produced by letting the escaping pairs undergo 
``secondary Penrose Compton scattering'' (SPCS) with low energy 
($0.03$~MeV) radially infalling equatorial 
accretion disk photons ($\equiv f_3$).  

Tables~1 and~2 give
model parameters corresponding to some of 
the numbers on Figure~1$a$ [see Williams (2003) for other numbers].
On these tables the parameters are defined as follows:
$r$ is the scattering radius; $E_e$ is the target
electron energy for PCS; $(E_\mp)_{\rm peak}$ is the energy
value where most of the PPP (\gggg) electrons,  used
as targets for the SPCS, are created;
$\nu_{\rm ph}$ is the initial infalling incident
photon frequency;
$\nu_{\rm peak}$  and $L_{\rm peak}$
correspond to the points (solid squares or dots superimposed 
on the small-dotted or dashed curve, respectively)
which give the continuum
luminosity resulting from  several distributions of PCS or SPCS
events (each distribution has 2000 scattering
events); $L_{\rm obs}$ is the observed luminosity at
$\nu_{\rm peak}$ (the average
frequency of the interval $\DD\nu$ where most of PCS or SPCS
photons are emitted per 2000 scattering events).
Each distribution of 2000
infalling photons have monochromatic energies normalized
to the  power-law distribution 
 for 3C~273 based on observations.  The $f_n$ values given
in the brackets are values used to fit the general model spectra
to agree with specific 
observations.
Overall, to produce the calculated Penrose
luminosity spectra of Figure~1$a$,
 74,000  infalling
photon scattering events are used.

Thus, as one can see from Figure~1$a$, the Penrose-Williams
 mechanism
can generate the necessary luminosity observed, and the three
model calculated regions of emission [due to PCS by equatorially 
confined targets 
(curve passing through nos.~$1-7$),
by nonequatorially confined targets that cross  the 
equatorial plane (curve passing up from 6 through nos.~$8-13$), and 
PPP (\gggg) (curve passing through nos.~$14-25$)] are consistent with
the three major regions of emission in all quasars and AGNs.   
Moreover, taking into consideration the characteristic voids, discussed
earlier, proposed to be produced by the different phases of the 
accretion disk, the lack of participating particles for PCS would  
cause a void between points~5 and~7 
($\sim 261~{\rm keV}-1.6$~MeV) on Figure~1$a$, suggesting a transitional 
energy regime between 
a thin disk ($\sim 3.5$~keV) and ion corona ($\sim 50$~keV).  
Comparing the observed spectrum of 3C~273 to the model calculated 
spectrum, it appears that this 
quasar has a similar accretion disk structure.  This  suggests
that a second void should  occur between $\sim 7~{\rm MeV}<E<107$~MeV,
as it does, agreeing strikingly well with observations
between the energies of points $13$ and $17$. Further, comparing 
Figures~8$a$ and~8$b$ of Williams
(2003), where Figure~8$b$ is the same as Figure~1$a$ of this present 
manuscript, Figure~8$a$ (quasar 3C 379) does not appear to have an 
appreciable inner region 
thin disk to effectively populate and depopulate equatorially confined 
target electron orbits
for PCS, which would give energies up to $\sim 262$~keV, like that 
of Figure~8$b$.  The lack of populating the equatorially confined 
target orbits  suggests that $Q_e\neq 0$ for the disk
electrons, implying the presence of an ion corona or ADAF,  
and leads to speculation that perhaps the equatorially confined ($Q_e=0$) 
particles were 
``lost'' in a 
prior thermal, yet cooler, instability phase.  This
interpretation for the accretion is consistent 
with observations of 3C~279, displayed
in Figure~8$a$: 3C 279
appears to have an 
effective radiating 
ion corona up to $\sim 40$~keV, 
similar to the case of item~($b$), in 
identifying the voids, but it lacks appreciable 
evidence of an inner region thin disk, at least at this time of 
observation---i.e., since 3C 279 is classified as an
optically violently variable (OVV) quasar (see Williams 2003). 
Such ADAF phase, presumed for 3C 279, could produce particles
to populate 
the nonequatorially confined target 
electron orbits for PCS and subsequent PPP (\gggg;  see 
items~2.$a$$-$2.$c$ above)
and could satisfy the conditions to produce Eilek's high energy
particles (see items~3.$a$ and~3.$b$ above; compare also Table~1 of
Williams 2003), giving rise to a model calculated spectrum 
consistent with the observed spectrum of 3C 279 (compare
Fig.~8$a$; see Williams 2003
for further details; see also \S~\ref{sec:3c273}).  

The observed spectra of microquasars (or galactic black
holes), in general, appear not to have PCS emission by the nonequatorially 
confined target electrons, neither the highest energy $\gamma$-ray 
emission due to PPP (\gggg), suggesting that these sources may not 
have an ion corona (or ADAF),  which 
would be need to populate the orbits to generate such emission, at
least in the highest energy regime (compare Figure~1$a$).
General calculated
spectra resulting from a self-consistent thin disk Penrose process
model  for stellar mass black holes ($\sim 30 M_\odot$) 
appear like a scaled-down Figure~1$a$
(with photon luminosity $\sim 10^{38-42}~{\rm erg\,s^{-1}}$ for total 
energy range $\sim 1$~keV$-$8~MeV), without any 
appreciable  curve labeled between 
points 6$-$13 (Williams \& Hjellming 2002); 
see \S~\ref{sec:cyg}.  This is consistent 
with observations of galactic black holes (Liang 1998).

\subsection{The Gravitomagnetic Field and Intrinsically Asymmetrical 
Polar Jets}
\label{sec:asymmetry}

The gravitomagnetic (GM) force field is the gravitational analog
of a magnetic field.
It is the additional
gravitational force that a rotating mass
produces on a test particle.  The
GM force 
is produced by the gradient of
${\vec \beta_{_{\rm GM}}}= -\omega{\bf  e}_\Phi$,
where $\omega$ is the frame dragging velocity (Bardeen et al. 1972)
and ${\vec \beta_{_{\rm GM}}}$ is the GM potential
(Thorne,  Price, \&  Macdonald 1986).
Analysis of the equations governing the trajectories of the 
Penrose process particles shows that 
the GM force, which acts
proportional to the momentum of a particle,  alters the
incoming and outgoing momentum parameters  of the incident 
and scattered particles, resulting in
asymmetrical polar distributions, and thus, appearing to break
the reflection symmetry of the Kerr metric, above and below the
equatorial plane (Williams 2004, 2002, 2003, 1999).\footnote{This effect
of apparent symmetry breaking has recently been confirmed by 
Bini et al. (2003): 
from a geometrical 
analysis of GM influence on spiraling Wilkins' (1972) nonequatorially 
confined test 
particle orbits 
in the Kerr-Taub-NUT spacetime.}  Effects
of the GM force acting on the PPP (\gggg) process can be discerned from 
comparing Figures~1$d$ and~1$e$. 
When half of the 2000 target photons are allowed to
have  initial polar coordinate momentum $(P_{\gamma 2})_\Th>0$ 
and the other half  $(P_{\gamma 2})_\Th<0$, 
of equal absolute values, with increasing $E_{\gamma 2}$, 
the $e^-e^+$ ``jet ($+{\bf  e}_\Th$) 
to counter-jet ($-{\bf  e}_\Th$)'' ratio $\ep_\mp$  
achieves a maximum $\sim 3:1$, favoring $(P_\mp)_\Th>0$
(Williams 2002), as seen in Figure~1$e$
(compare Fig.~1$d$).
The corresponding polar angles of escape for cases of Figures~1$d$
and~1$e$ are displayed in Figures~2$a$ and 2$b$, respectively.
Polar coordinate momentum distributions, $(P_{\rm ph}^\prime)_\Th$, 
for escaping PCS photons are displayed in Figure~3, where the primes
indicate final conditions.   The corresponding polar angles of escape
for the cases of Figure~3 are given in Figure~4.
Notice the effects of
the GM force field causing the (photon jet to counter-jet) ratio 
$\ep_{\rm ph}$
to vary from nearly symmetric 
to asymmetric
for the different cases shown.  Of these cases the largest ratio 
achieved is $\sim 5:1$ (Figs.~3$c$ and~4$c$).
The direct cause of the asymmetry in the polar direction appears
to be due to the severe inertial frame dragging 
in the ergosphere in which
the GM field lines are spacetime dragged in the direction 
that the black hole
is rotating [see Williams (2002) for details; see also Williams
2004].  The resulting GM 
force acting on the 
particles produces the asymmetry.
 
In most cases, the distribution favors the  $+{\bf e}_\Th$
direction (see Figs.~3 and~5$d$); however, at particularly  low energies,
 the asymmetry appears to reverse.
For example, in the case of PPP (\gggg) at 
the low initial energies
$E_{\gamma 1} =3.5$~keV and $E_{\gamma 2}\simeq 3.4$~MeV for the infalling
and orbiting photons, respectively, producing escaping 
$e^-e^+$ pairs with 
energies peak around $E_\mp\sim 1.5$~MeV, $\ep_\mp=700/615\simeq 1.14$ 
per 2000 events (Fig.~5$a$),  and after undergoing SPCS (Williams 2003) 
per 2000 infalling disk
photons ($E_{\rm ph}=3.5$~keV), the asymmetry in the final photon polar 
distribution, for the SPCS, is reverse, with the inverse of the number of 
particles scattered in the positive polar direction to that in the negative
direction 
$[\ep_{\mp({\rm ph})}]^{-1}=402/165\simeq 2.44$, favoring the 
$-{\bf  e}_\Th$ direction (Fig.~5$b$).
This would make the $-{\bf  e}_\Th$ jet appear more energetic 
and, thus, brighter, since the PPP $e^-e^+$ polar jets, in this case,
are nearly symmetrical, as can be seen in Fig.~5$a$. 
Such behavior is consistent with Hjellming and 
Rupen's  (1995) observations of 
GRO J1655-40. These authors concluded that the jets themselves 
must be intrinsically asymmetric, and the sense of the asymmetry
must change from event to event.  Moreover, they found that 
the jets lie almost in the
plane of the sky, so relativistic beaming cannot explain the
observed brightness ratios.  [Note, the potential for 
``jet reversal''
due to the GM force field can be seen in eq.~(47) of Williams (2002) 
and eq.~(8) of Williams (2004): occurring 
 for particle distributions
with  relatively large $P_r^\prime>0$
 and/or relatively small 
$P_\Phi^\prime\equiv L^\prime$ (corresponding to 
small $E^\prime$).]  Also, the jet space velocity 
Lorentz factor 
found by these 
authors ($\Gamma=[1-(v/c)^2]^{-1/2}\approx 2.5\Ra E\sim 1.3$~MeV) 
is consistent with the target electron energy, of the  
SPCS, we have found here, displaying 
the jet reversal (compare $E_\mp$ stated above and Figs.~5$a$ and~5$b$), 
where we are assuming
that the bulk velocity of a ``blob'' is $\sim\langle v_\mp\rangle\equiv$
``average'' space velocity of the individual PPP electrons per
bulk distribution, i.e., assuming $ \langle\gamma_\mp\rangle\sim\Gamma$, 
valid at least in the case of the small scale, fast varying galactic
black holes.  The model calculated space velocities 
of the
PPP $e^-e^+$ jet particles for initial energies consistent 
with observed
microquasars (with $M=30M_\odot$) are in the range 
$v_\mp\sim 0.83-0.98c$, for $E_\mp\sim 0.9-2.4$~MeV, implying 
$\gamma_\mp\sim 1.8-4.7$, respectively; compare Fig.~1$f$.  
Thus, the consistency of apparent jet reversal, of
these Penrose processes, with observations,
gives more compelling 
evidence that it is probably the 
Penrose-Williams mechanism at work, close to the event horizon,
within $r_{\rm ms}$,
extracting rotational gravitational energy momentum: in the form
of a particle flux, as opposed to the so-called BZ-type models,
proposed to extract energy and momentum: in the form of electromagnetic 
Poynting flux and Alfv\'en waves, respectively (with the major
problem
still existing of converting to the  necessary 
particle flux to fuel the observed jets). 

Note, a specific possible scenario for the jet reversal in the 
case of a $30M_\odot$ microquasar, similar to that of GRO J1655-40 
(Hjellming \& Rupen 1995), for a classical thin relativistic 
accretion disk (Novikov \& Thorne 1973), is the following: as 
secular density and thermal 
 instabilities begin occurring in the inner region of a
time dependent accretion disk---commonly  
referred to as the ``Lightman instabilities'' 
(Lightman 1974a, 1974b; Williams  1995, 2003), $kT_e$ increases to a 
``reasonable'' maximum $\sim 30$~keV, being consistent
with observations.  
The infalling disk 
particle electrons with energies 
 $17~{\rm keV}\,{<\atop\sim}\, E\,{<\atop\sim}\, 33~{\rm keV}$,
and satisfying  conditions
for a turning point to exist at  specific radii (Williams  1995) between 
$r_{\rm mb}\,{<\atop\sim}\,r\,{<\atop\sim}\,r_{\rm ms}$, respectively,
will be gravitational blueshifted according to the blueshift factor:
$32\,{>\atop\sim}\,e^{-\nu}\,{>\atop\sim}\,10.7$, respectively 
(recall discussion in \S~\ref{sec:model}, first paragraph), 
populating the 
 equatorially confined  ($Q_e\simeq 0$) target electron orbits with 
$0.5388~{\rm MeV}{>\atop\sim}\,E_e{>\atop\sim}\, 0.3486$~MeV, 
respectively, for PCS (Williams 1995).  
This appears to be the catalyst to
``turn on'' the self-consistent Penrose-Williams mechanism.  
[Note, the above reasonable maximum energy means before the critical 
surface density $\Sig_{\rm crit}(r)$
is reached (Lightman 1974a, 1974b),
which causes the ion coronal/torus two-temperature
phase to set in, or before the inner ``hot'' region 
(Novikov \& Thorne 1973)
extends to ${>\atop\sim}\, 90M$ (Eardley \& Lightman 1975),
for $0.001\leq\alpha\leq 0.1$,
$y=0.6$, and
$1\times 10^{-8}\,{<\atop\sim}\,{\dot M}/M_\odot\,{\rm yr^{-1}}\,
{<\atop\sim}\, 3\times 10^{-8}$,  where $\alpha$ is the viscosity
parameter, $y$ the Kompaneets parameter, and $\dot M$
the sub-Eddington accretion rate; see Williams (2003).] 
The subsequent escaping PCS X-ray emission 
becomes more and more asymmetric, 
favoring the $+{\bf e}_\Th$ direction,
as the infalling  initial photon energy is increased, say due to 
disk instabilities (compare Figs.~3$a$ and 3$b$).  As PCS of infalling 
disk photons ($E_{\rm ph}=3.5$~keV)
depopulates the equatorially confined
target electron orbits, some of the photons 
with $(P_{\rm ph}^\prime)_r<0$, $12~{\rm keV}\,{<\atop\sim}\,
E_{\rm ph}^\prime\,{<\atop\sim}\,111~{\rm keV}$, and 
$Q_{\rm ph}^\prime\,{<\atop\sim} 0.07M^2m_e^2$,
after being gravitationally blueshifted by factor $e^{-\nu}\sim 52$, 
satisfy  conditions to have a 
turning point at  the 
photon orbit (Williams 1995, 2002), populating, and thus supplying target 
photons for PPP (\gggg) in the range of $0.6~{\rm MeV}\,{<\atop\sim}\,
E_{\gamma 2}\equiv E_{\rm ph}^{\prime\prime}\,{<\atop\sim}\,5.8~{\rm MeV}$,
respectively, for the range of $E_{\rm ph}^\prime$ above,
where $E_{\rm ph}^{\prime\prime}=e^{-\nu}E_{\rm ph}^\prime$.
For $E_{\gamma 2}\sim 3.4$~MeV,
$Q_{\gamma 2}\sim 0.016\, M^2m_e^2$, as given by the analytical  
derived expressions of the conserved energy $E$ and angular
momentum $L$ of nonequatorially
confined particle trajectories  (see Williams 1995, 2002, 2004),
 the subsequent PPP (\gggg) with infalling disk photons 
$E_{\gamma 1}=E_{\rm ph}$ 
(assuming negligible electrons are left in the 
equatorially confined orbits between
$r_{\rm ms}$, $r_{\rm mb}$) will produce slightly asymmetrical 
jets (favoring the $+{\bf e}_\Th$ direction;
compare Fig.~5$a$).  The total energetics due to PCS and 
PPP (\gggg) at this phase
will favor $+{\bf  e}_\Th$, therefore, producing a ``brighter'' 
jet in this polar 
direction.  However, when some of these PPP electrons subsequently 
interact with infalling disk photons through 
SPCS, the final emitted
escaping photon jets undergo apparent reversal 
(favoring $-{\bf  e}_\Th$; compare Figs.~5$b$ and~5$c$):
thus, the total energetics will now favor the $-{\bf e}_\Th$
direction.
Compare Figures~3$a$, 3$b$, and 5$a$$-$5$c$, considering 
 the observed time delays between outbursts (Hjellming \& Rupen 1995; 
Eikenberry et al. 1999a) and those expected 
between the different Penrose
processes: PCS, PPP (\gggg), SPCS;  and the
 synchrotron emission
($\sim 1.4-8.5$~GHz) by the escaping PPP electrons
(of Fig.~5$a$): due to, perhaps, their expected 
intrinsic magnetic field (or an external accretion disk 
magnetic field), 
according to $\nu_{\rm syn}\sim 4\times 10^6 \gamma_e^2 B $ 
(Burbidge, Jones, \& O'Dell 1974), for 
$B\sim 10^2$~G (Williams 2003; this assumed value, although
consistent with observations in many instances,
 needs further 
investigation).  
In addition, some of
the PPP electrons (of Fig.~5$a$) will be created  with 
$E_\mp{\sim} 1.35$~MeV, 
$Q_\mp\sim 0.076M^2m_e^2$, $(P_\mp)_r \sim 7.2 m_e$, 
and $L_\mp\sim 5.6 Mm_e$ (recall that $G=c=1$),
satisfying the condition to have
a turning point at the 
iso-energy orbit $E_{\rm orb}$ (circular orbit of equal energy 
at constant radius $r=r_{\rm orb}$; see Williams 1995, 2004),
with $E_\mp=E_{\rm orb}$ and $L_\mp>L_{\rm orb}$ at  radii
$r_{\rm orb}\sim r_{\rm mb}$ (the last bound orbit for a 
material particle,
deep within the ergosphere), before escaping to
infinity along vortical orbits (\S~\ref{sec:collimation}), 
satisfying (Williams 1995)
\begin{equation}
0<{Q_\mp\over E_\mp^2}<{Q_{\rm orb}\over E_{\rm orb}^2},
\end{equation}
or $Q_\mp<Q_{\rm orb}$, implying no turning point in $(P_\mp)_\Th$, 
i.e., $(P_\mp)_\Th\nrightarrow 0$, yet $(P_\mp)_r\rightarrow 0$.  
Note, this satisfying of the condition 
to have a turning point at $r_{\rm orb}\sim r_{\rm mb}$, before
escaping to infinity along vortical trajectories,
 is also true for the 
supermassive KBH (Williams 2004).
Observations of GRO J1655-40
(Hjellming \& Rupen 1995) suggest
that after the jet outbursts: due to Lightman instabilities, inner
region disk depletion, Penrose processes, and  plunging
orbit (Bardeen et al. 1972) population-depopulation processes, 
 the disk settles back down to its low, ``initial'' state, to prepare 
 once again to repeat the total {\it disk instability-Penrose 
emission cycle}, as 
described above, indefinitely (i.e., as long as there exists 
available matter to
accrete).  Moreover, the disk instabilities are expected to change the 
accretion rate, thereby
causing the Penrose processes to vary.

So, in conclusion of this section, it appears that once the initial 
requirement has been met:
of populating the equatorially
confined target electron bound, unstable orbits, inside
the ergosphere, between $r_{\rm mb}\la r\la r_{\rm ms}$ 
[at $\sim r_{\rm mb}$ for
maximum PCS energy extraction (Williams 1995)], 
the KBH operates as a self-consistent system, emitting $e^-e^+$ 
and photon jets, relying only on the 
accretion disk to supply the incident infalling photons, and to
populate the initial equatorially confined electron target orbits 
[i.e., due to disk instabilities (Kafatos \& Leiter 1979)]---indicating 
the beginning of the ``cycle.''  And within this cycle for particularly
low particle initial energies, the GM field can cause the jet brightness
asymmetry to reverse. [Note, see Williams (2002)  for a
complete description of the relations between the GM field
and the space momenta displayed in the figures
shown here.]
In addition, in the case
of quasars-type AGNs (Williams 1995, 2003), 
it appears that an ADAF is needed to populate
the relatively high energy 
nonequatorially confined target electron orbits for PCS;  
and to populate the highest energy photons at the photon 
orbit for PPP (\gggg),
yielding maximum escaping energies 
$E_\mp\sim 54$~GeV 
(as discussed in \S~\ref{sec:energy-luminosity}).

\subsection{The Vortical Orbits and Intrinsically Collimated Polar Jets}
\label{sec:collimation}

It is found that the Penrose scattered
particles escape along vortical trajectories 
collimated about the
polar axis (Williams 1995, 2000, 2003, 2004).
These distributions  are fluxes of coil-like
trajectories of relativistic jet-type particles, 
 escaping out from the equatorial plane at the 
scattering radius $r_{\rm ph}\la r\la r_{\rm ms}$, concentric
 the polar axis.
The highest energy particles
have the largest $P_\Phi^\prime$ values (compare Fig.~1$c$; 
compare also Figs.~3$b$ and~4$b$ of Williams 1995).  
Note, $P_r^\prime$ is negative
(inward toward the polar axis) for many of the PCS photons
(Williams 2002), 
and positive 
for all of the $e^-e^+$ pairs (compare Fig.~1$b$).
The helical angle of escape  $(\dd_i)_{\rm esc}=
\vert 90^\circ -\Theta^\prime\vert$, of
particle type $i$, relative to the equatorial plane,
for the highest energy scattered
particles ranges from $(\dd_{\rm ph})_{\rm esc}\simeq 1^\circ$ to
$30^\circ$ for PCS (compare Fig.~4) and
$(\dd_\mp)_{\rm esc}\sim 0^\circ .5$ to $20^\circ$
for the $e^-e^+$ pairs (compare Fig.~2);
compare also Figs.~6, 7, and~9 of Williams (2002).  The above
characteristics of the escaping particles, along with
their $\vert P_\Theta^\prime\vert$ values (compare Figs.~1$d$, 1$e$,
3, and~5), imply
 strong collimation about the polar axis,
giving rise to relativistic jets with
particle velocities up to
$\sim c$  (compare Fig.~1$f$).  Note, 
such vortical trajectories and 
collimation are consistent with the findings of de~Felice et~al.
(de Felice \& Curir 1992, de Felice \& Carlotto 1997,
de Felice \& Zanotti 2000),
 from spacetime geometrical studies of  general particle geodesics
in a Kerr metric.  Moreover, the GM force field, discussed in the
last section, responsible for the inertial frame dragging and the 
asymmetrical jets, also serves to boost the jets into opposite polar 
directions (Williams 2002).

\subsection{Agreement with Observations}
\label{sec:observation}

\subsubsection{Quasars 3C 273 and 3C 279}
\label{sec:3c273}

In addition to statements made  in 
\S~\ref{sec:energy-luminosity} 
concerning the model calculated spectra of 3C~273 and 3C~279, 
below I summarize some of the important 
features resulting from application of 
 the Penrose-Williams  mechanism 
 to  observations of both 3C~273 and 3C~279 (see Fig.~8$a$ of 
Williams 2003).
The observed spectra of both these sources
can be explained by these Penrose processes 
and the 
assumed accretion model: specified  in \S~\ref{sec:model}
[see Williams (1995, 2003) for further details].  
As we can see from Figure~1$a$,
there is a striking similarity between the energy range of
the observed spectrum of 3C~273 and the model spectra produced by 
these Penrose
processes.   Upon comparing the 
spectra of radio-loud quasars 3C~273 and 3C~279,
based on these Penrose processes, 
we find the following (Williams 2003): 
 the shape of the observed spectrum of 3C~273
looks like the ``enhanced'' (i.e., the highest 
observed energetic state)
 spectrum of 3C~279, except for the higher
luminosities in 3C~279 and the radio tail in  3C~273.
The higher luminosity and the apparent lack of a
radio tail in 3C~279 is
probably, largely, due to the radiation of 3C~279 being beamed more
in the direction of the observer than the radiation of 3C~273.
Therefore, the spectrum of 3C~279 has been Doppler blueshifted to an 
{\it observed\/} higher
energy interval; and the {\it apparent\/} luminosity has been increased.
This is consistent with radio observations which detect more
superluminal motion (or relativistic beaming near the line of sight
of the observer)
 in 3C~279 than in 3C~273 (Porcas 1987).  On the other hand,
it seems that 3C~273 has a ``hotter'' inner accretion disk and 
is in a predominantly bimodal quasi-stable state: appearing to be in 
the most effective or ``extreme'' thin disk/ion corona state 
as opposed to 3C~279: which appears to
oscillates in a highly
variable
fashion between the thin disk and ion corona phases---for this reason
3C~279 is classified as an OVV quasar.
The hotter state of the accretion disk (ion corona),
which is heated
by a runaway thermal instability (Shapiro, Lightman, \& Eardley 1976),  
would result in
enhanced Penrose processes [PCS and PPP (\gggg)],
and enhanced synchrotron
radiation due to the presence of more relativistic electrons,
particularly if Eilek's (Eilek 1980; Eilek \& Kafatos 1983) 
particle reactions 
($pp\ra \pi^0\ra\gamma\gamma\ra e^-e^+$) 
occur, hence contributing  to the prominent observed 
radio tail of 3C~273.  
This ion corona/ADAF state, existing in conjunction with the 
thin disk,  appears to be the case always in the 
continuum emission of 3C~273
and sometimes in the emission spectrum of 3C~279, with 3C~279 not
quite achieving the full ``hot'' 
ion corona/ADAF status of 3C~273 (Williams 2003),
neither achieving the full ``cool'' $E<3.5$~keV thin disk phase,
where more Penrose processes would occur to liberate trapped energy.
Thus in summary, the
differences in the spectra of 3C~279 and 3C~273 are probably
due to (1) the more beaming effect in 3C~279, and (2)
the predominantly extreme hot, cool 
phases of 3C~273.  Now, again, based on the ``characteristic voids''
discussed in \S~\ref{sec:energy-luminosity}, 3C~279 appears to 
be similar to case (b) and 3C~273 to that of case (c). 
Compare models $1-3$ and $6$ on Table~1 of Williams (2003): models $1-3$
are similar to 3C~273, and model $6$ is similar to 3C~279.

\subsubsection{Seyfert 1 Galaxy MCG---6-30-15}
\label{sec:MCG}

Recent observations of the
bright Seyfert~1 galaxy
MCG---6-30-15 [particularly of the broad Fe K$\alpha$ emission
line at $\sim 6$~keV, believed to be originating
from the inner accretion
disk plasma (Wilms et al. 2001)],
and other such type AGNs,
are consistent with these
 model calculations.  A qualitative model calculated scenario
to explain the observed spectral
observations of MCG---6-30-15, by these Penrose processes, is
as follows.
If we assume that the plunging orbits of the target electron, 
inside the ergosphere,
have been populated by accretion disk instabilities
(as described in \S~\ref{sec:asymmetry}),
 self-consistent  computer simulations of these Penrose
processes
 consistent with MCG---6-30-15  have model
parameters for  radial infalling photons ($E_{\rm ph}=2$~keV)
from a thin
disk (Novikov \& Thorne 1973), that either undergo PCS by
equatorially 
confined orbiting
target electrons ($E_e\simeq 0.539$~MeV) at $r_{\rm mb}$, or
PPP ($\gamma \gamma \longrightarrow e^- e^+$\/) at $r_{\rm ph}$.
The 
 energies (due to frame dragging) attained by
the $\sim 31\%$ up to $83\%$ escaping particles, returning to the
disk to be reprocessed and/or escaping
to infinity, are the following:  For PCS photons,
$E_{\rm ph}^\prime\sim 5.2-175$~keV
for equatorially  confined orbiting
target electrons, 
with relative incoming and
outgoing photon luminosities
$(L_\gamma)_{\rm out}\sim 0.014-11\,(L_\gamma)_{\rm in}$,
respectively, where 
$(L_\gamma)_{\rm in}\sim 2.5\times 10^{42}~{\rm erg\,s^{-1}}$. 
And for the relativistic PPP
electrons (with $E_{\gamma 1}\equiv E_{\rm ph}$ and
$E_{\gamma 2}\simeq 4.8$~MeV),
$E\mp\sim 2.4$~MeV [consistent with synchrotron radiation
into the radio regime for $B\sim 10^{2}$~G, and inverse Compton
scattering (SPCS of disk photons) into the
X-ray/soft $\gamma$-ray  regime---with relative incoming and
outgoing photon luminosities $(L_\gamma)_{\rm out}\sim 0.006-2.7\,
(L_\gamma)_{\rm in}$, for $M\sim 10^8M_\odot$,
at $\sim 71~{\rm keV}-1.3$~MeV, respectively],
suggesting  relatively weak, less powerful
and less prominent radio
jets, i.e., a radio quiet AGN, like a Seyfert galaxy 
(compare Figs.~1$a$, 5$a$ and 5$b$ for similarities and
dissimilarities).  
Note, for self-consistency, $E_{\gamma 2}$ is assumed based 
on prior PCS
photons with $(P_{\rm ph}^\prime)_r<0$ that satisfy conditions
for the existence of a turning point at the photon orbit (Williams 2002). 
Note also that, at these low energies for $E_{\gamma 1}$ and 
$E_{\gamma 2}$, the SPCS polar jets appear to ``flip,'' undergoing
brightness jet reversal (as discussed in \S~\ref{sec:asymmetry}), 
differing by
a factor $\sim 10.6$, in particle numbers,
 favoring $-\unitt$ (compare Fig.~3 and 
Figs.~5$b-5d$), whereas  the initial PPP 
target electron  
polar jets,  differ by a factor  
$\sim 2$ favoring $+\unitt$ (compare Fig.~5$a$).  
The PCS photon distribution in the range such as 
 $E_{\rm ph}^\prime$ above, 
emitted from $r_{\rm mb}\la r\la
r_{\rm ms}$, with the highest energy photons concentrated in the 
equatorial plane (compare Fig.~4), is expected to be consistent with the 
observed extremely steep emissivity profile $\bb\sim 4.3-5.0$ of 
Wilms et al. (2001), indicating  that most of the Fe K$\alpha$ line    
emission originates from the inner region of a relativistic 
accretion disk.  Specific details of the emissivity
$\eps(r)\propto r^{-\bb}$ of these Penrose processes, particularly of
the PCS, will be presented in a future paper by the author.

\subsubsection{Radio Galaxy M87}
\label{sec:M87}

Recent radio observations
of active galaxy M87 (Junor et al. 1999)
suggest that electromagnetic
collimation becomes important at radii ${>\atop\sim}\, 30-100 r_g$,
wherein the initial ``open angle'' of the jet $\sim 60^\circ$
(at radii $< 30 r_g$) is
made smaller to $\sim 30^\circ$ by the electromagnetic field,
where  $r_g=2M$ ($=r_0$, the radius of the ergosphere at the equator).
This is consistent with the Penrose mechanism providing (in addition
to the relativistic particles) the initial
collimation  at radii
($< 30 r_g$), i.e.,  closer
to the black hole.
Since M87 is a giant elliptical galaxy, this
could mean that its geometric configuration is
possibly helping to
maintain the
initial collimation by the black hole (Williams 2003): which begins at
$r<r_g$, and must extend out to at least
$\sim 30r_g $---i.e., until, it appears, electromagnetic
collimation takes over.  However, before one can say for certain
of the electromagnetic processes occurring,
 a time dependent MHD evolution of the Penrose escaping particle
plasma must be performed (presently under investigation by the
author).  It should not be ruled out that the
intrinsic collimation due to the black hole, of the escaping
relativistic plasma: and any associated ``dynamo''
 generated magnetic field,
may be sufficient to maintain collimation.

Further, concerning M87, its observed spectrum in general can be
explain by the Penrose mechanism presented in this paper.
Some observational properties of M87 are the following (Eilek 1997):
$L_{\rm jet}\sim 10^{43-44}{\rm erg\,s^{-1}}$; striking
comparisons of radio (Very Large Array) and optical (Hubble 
Space Telescope) images of the jet;
optical  and possibly X-ray emission believed to be of
synchrotron origin; and more recently, the 
 mid-IR observations (Perlman et al. 2001) showing that the nuclear IR 
emission is entirely consistent with
synchrotron radiation, and there is no evidence for 
thermal emission from a dusty nuclear torus.
Based on these properties
the following scenario can
be devised according to the Penrose-Williams mechanism.
The jet is no doubt beamed, since observed superluminal
motions give
apparent velocities up to $\sim 6c$, implying
line-of-sight angle $\theta_s\sim 10^\circ-19^\circ$,
bulk Lorentz factor $\Gamma\sim 6-40$,
jet Doppler factor $\dd_*\sim 5.7-0.5$, and jet brightness boost
$\dd_*^3\sim 190-0.1$, respectively (Biretta, Sparks, \& 
Macchetto 1999; see also Williams 2003).  
M87 is probably an evolve blazar-type
AGN (OVV quasar
and BL Lac object).  Its luminosity spectrum (although less powerful,
less energetic)
most likely resembles that of 3C~279 
(\S~\ref{sec:3c273}; see also Williams 2003).   
The most noticeable
change in the spectrum from
times past is probably the lack of high energy $\gamma$-rays: due
to the lack of the availability of infalling low energy
(soft X-ray) disk photons, or the lack of
high energy PPP (\gggg) electrons,
to  undergo effective SPCS, that would result in escaping
trajectories for the scattered $\gamma$-rays.
Since the jet of M87 is still seen prominently in the
radio/optical/X-ray, an optically thin hot ion  torus, PCS,
PPP (\gggg),
and subsequently synchrotron radiation of the PPP electrons
(particularly into the optical: implying $E_\mp \sim 177-558$~MeV
for $B\sim 10^{3-2}$~G, respectively),
are consistent with the observations.  The parenthetical
statement above suggests that the
magnetic field producing the  synchrotron
radiation may be that
of the escaping Penrose plasma rather than that of the
popular proposed large
scale dipolar-like field of the accretion disk (since large-scale, 
large-strength dipolar accretion disk fields are in practice 
difficult to create); this however requires an
investigation.  Moreover, besides coming from the inner region
of a relativistic thin disk (Novikov \& Thorne 1973),
 there are two possibilities
for producing the
observed soft X-ray emission, within the confinements of the
Penrose-Williams mechanism: (1) a synchrotron origin
requires electron energies $\sim 17$~GeV
($\gamma_e\sim 3\times 10^4$) for $\sim 2$~keV emission at
$B\sim 10^{2}$~G, and could very well be produced
by the PPP (\gggg), at least for ultrarelativistic $e^-e^+$
pairs up to $\sim 54$~GeV (Williams 2003) for $B$ as low as 
$\sim 10$~G; 
 and (2) the jet is
beamed, and
self-Compton scattering
of lower energy radio and IR synchrotron photons
by the escaping, intrinsically polar collimated
PPP electrons is occurring:
the observed energies of the
inverse/self-Compton scattered photons are  blueshifted due
to Doppler boosting into the optical and X-ray regimes,
respectively, according to
 $E_{\rm Comp}\simeq 0.5\gamma_e^2 h\nu$ (Dermer, Schlickeiser, 
\&  Mastichiadis 1992)
for $\Ga=6$, $\theta_s=10^\circ$.
Now, both items above could equally occur, more or less;
however,
 since superluminal motion appears to be important in
M87,  item~(2) is most likely the dominant, somewhat ruling out the
other item.  If this dominance is true, then
the energies of the jet electrons need only be as high as
$E_\mp\sim 20-150$~MeV for $B\sim 10^{2}$~G.  
This is consistent with the Penrose processes
described here, in the presence of a thin disk/ion corona  
accretion---without
the need of Eilek's $\pi^0$ decays to populate the photon
orbit  (see \S~\ref{sec:energy-luminosity}, items $2.a-2.c$).
Note,  such ion coronas or tori are poor radiators, 
and expected
to be of relatively low density, with $30~{\rm keV}\la kT_e\la 50$~keV; 
this may account for the 
lack of evidence for 
an inner ``dusty'' torus emitting thermal radiation 
in the mid-IR observations by Perlman et al.
(2001).  However, it still seems unlikely that such a low
electron energy and particle density ion torus 
(or ADAF)
can be jet fuel for the BZ-type  models near the event
horizon, inside the ergosphere, 
 as required by such models (Blandford \& Begelman 1999).  
[See Williams (2003) for 
a complete description of the accretion 
disk model consistent with these
Penrose processes and observations.] Nevertheless, such BZ-type models
(e.g., Punsly 1991; Koide et al. 2000) might be important at 
$r>30 r_g$, as suggested by observations (Junor et al. 1999),
particularly if the Penrose-Williams particles are used as fuel.

\subsubsection{Galactic Black Hole X-Ray  Source Cygnus X-1}
\label{sec:cyg}

The Penrose-Williams model presented here applies
to all  mass size KBHs,  with the stellar mass black hole appearing
as a scaled-down supermassive hole.  When the parameters are expressed
in gravitational units ($c=G=1$), the Penrose process emission
energy-momentum spectra ($P_r$ vs.~$E$; $P_\Th$ vs.~$E$; 
$P_\Phi$ vs.~$E$) over
the range of masses are approximately identical.  The luminosity
spectra of these Penrose processes for the different masses, 
in general, span over a 
range $\sim 10^{38}-10^{52}~{\rm erg \,s^{-1}}$ (compare Figs.~1$a$ 
and~6).
  In general, the differences of the Penrose process 
output luminosities
between supermassive KBHs and ``micro-massive''  KBHs are determined 
by the bolometric luminosity of the incoming photons 
(Eilek 1980; Williams 2003), directly dependent on the accretion rate,
which is governed by the surrounding accretion disk
 environment.
For example, the
observations of the classical stellar/galactic black-hole candidate
 Cygnus X-1 (Liang 1998) can be explained by these Penrose processes:
 Processes consistent with Cyg X-1 have model
parameters for radial infalling photons ($E_{\rm ph}=3.5$~keV)
from a thin
disk (Novikov \& Thorne 1973), that either undergo PCS by
equatorially  confined orbiting
target electrons ($E_e\simeq 0.539$~MeV) at $r_{\rm mb}$ or
PPP ($\gamma \gamma \longrightarrow e^- e^+$\/) 
at $r_{\rm ph}$.
The 
 energies (due to frame dragging) attained by
the $\sim 82\%$ up to $92\%$ escaping particles, returning to the
disk to be reprocessed and/or escaping
to infinity, are the following:  For the PCS photons,
$E_{\rm ph}^\prime\sim 12-250$~keV,
with relative incoming and outgoing
photon luminosities 
$(L_\gamma)_{\rm out}\sim 0.4- 130 \,(L_\gamma)_{\rm in}$,
respectively, where 
$(L_\gamma)_{\rm in}\sim 4\times 10^{38}~{\rm erg\,s^{-1}}$.
And for the relativistic PPP
electrons (with $E_{\gamma1 }\equiv E_{\rm ph}$ and
$E_{\gamma 2}\sim 5$~MeV),
$E\mp\sim 4$~MeV [consistent with synchrotron radiation
into the radio regime for $B\sim 10^{2}$~G, and inverse Compton
scattering  (SPCS of disk photons) into the hard
X-rays/soft $\gamma$-ray regime---with relative incoming and
outgoing photon luminosities  $(L_\gamma)_{\rm out}\sim 8-2000 \,
(L_\gamma)_{\rm in}$, for $M\sim 30M_\odot$, between
$\sim 100$~keV$-3$~MeV]; compare Figure~6;
see Williams \& Hjellming (2002)
for further details.  Note, for self-consistency, $E_{\gamma 2}$ is 
assumed based on prior PCS
photons with $(P_{\rm ph}^\prime)_r<0$ that satisfy conditions
for the existence of a turning point at the photon orbit 
(Williams 2002).  Note also that, as in the cases of GRO J1655-40 
(\S~\ref{sec:asymmetry}) and MCG---6-30-15 
(\S~\ref{sec:MCG}),  at these low energies 
for $E_{\gamma 1}$ and
$E_{\gamma 2}$, the SPCS polar jets undergo slight so-called jet 
reversal (as discussed in \S~\ref{sec:asymmetry}), differing by
a factor $\sim 1.4$ favoring $-\unitt$, whereas  the initial PPP
target electron
polar jets,  differ by a factor
$\sim 1.1$ favoring $+\unitt$
(compare Figs.~3$a$,  3$b$,
and~5).

In the above model for Cyg X-1, the PPP electron energy $E_\mp$ 
can  increase to ${>\atop\sim}\,  10$~MeV, as the infalling thin
disk photon energy for PCS by equatorially confined target electrons
is increased to $\sim 20-30$~keV (Williams \& Hjellming 2002),
say due to disk instabilities (compare
\S~\ref{sec:asymmetry}).  This appears to be the case for Cyg X-1 when
in its ``high'' state (McConnell et al. 1989), and to 
explain the persistent
power-law $\gamma$-ray tail up to $\sim 20$~MeV (McConnell et al. 
1994).

Now, concerning $\sim {\rm kHz}$ quasi-periodic 
oscillations (QPOs) observed in galactic black holes
 (Strohmayer 2001; Remillard et al. 
2002;
Abramowicz et al. 2002), 
such QPOs can be predicted from the Penrose
scattering processes described here.  The
``QPOs'' of, say, a given local distribution of neighboring target 
electrons, responsible for  PCS  
 into the X-ray/soft $\gamma$-ray regime, emitting from  geodesic 
orbits at radii between $r_{\rm ms}$ 
and $r_{\rm mb}$, can be obtained from (Bardeen et al. 1972) 
\begin{equation}
\Omega={{e^\nu v_\Phi}\over\sqrt{g_{\Phi\Phi}}}+\omega,
\label{eq:QPO}
\end{equation}
where $\Omega\equiv d\Phi/dt$ is the coordinate angular velocity of 
a circular orbit;  $e^\nu$ is the inverse of the blueshift factor
(see \S~\ref{sec:model}), commonly referred to as the ``redshift'' factor;
$v_\Phi$ is the orbital velocity in the azimuthal direction of the target 
particles relative to the LNRF, i.e.,
as measured by a general observer at rest relative to this frame
(Bardeen et al. 1972; see also
Williams 1995); 
$e^\psi=\sqrt{g_{\Phi\Phi}}$ is the radius of the circumference
about the axis of symmetry (Thorne et al. 1986).  So, with 
the  frame dragging
{\it angular} velocity given by   
$\ww=\ww(r,a=0.998M,M=30M_\odot,\Th=\pi/2)$, 
we find (from eq.~[\ref{eq:QPO}]) the predicted range to be between: 
  $\nu_{_{\rm QPO}}=(\WW_{_{\rm QPO}}/2\pi)\simeq 467$~Hz, 
$\simeq 506$~Hz at 
$r_{\rm ms}$, $r_{\rm mb}$, respectively,
corresponding to periods $\sim 2$~ms, as measured by 
an observer at infinity.  
Note, the counterpart QPOs for a supermassive ($10^8M_\odot$)
 KBH are $\sim 2\times 10^{-4}$~Hz; this relatively low frequency
is probably the 
reason these counterpart QPOs have yet to be detected in sources 
harboring such massive KBHs (see Miller et al. 2002 and references 
therein).   These calculations suggest that 
$\sim $~kHz QPOs may also be due to the inertial frame dragging of the 
nonequatorially confined target particles' orbital ``ring'' at
scattering radius $r$
(Williams 1995), particular of the nodes (points at which the orbit,
in going between negative and positive
latitudes, intersects the equatorial plane)---which happens to be 
where the most effective Penrose scattering processes would occur, 
as resulting emitting regions of neighboring
particles  sweep across the line of sight of the observer.
In this case,
the observed oscillation frequencies, given by $\omega$ 
(Bardeen et al. 1972), might be slightly smaller 
(between $\nu_{\rm QPO}\simeq 439$~Hz, $\simeq 494$~Hz at $r_{\rm ms}$, 
$r_{\rm mb}$, respectively) and appear
twice as fast as those given above or in pairs.\footnote{Similar 
effects have been independently suggested by Stella, Vietri,
and Morsink (1999),
concerning nodal precession, and by Cui, Zhang, and Chen (1998)
as evidence of frame dragging around spinning black holes.}
See Williams (2002) for a  discussion of the nonequatorially
confined spherical-like
orbits,  first proposed by Wilkins (1972).   
The above findings
are consistent with the QPOs proposed  to originate from orbits within
the radius of the marginal stable orbit $r_{\rm ms}$
(Zhang, Shrohmayer, \&
Swank 1997), and the suggestion that the energy 
distribution of the
energetic electrons must be oscillating at the QPO frequency
(Morgan, Remillard, \& Greiner 1997).

Note, in the above qualitative, yet self-consistent models,
 for the radio quiet Seyfert 
 galaxy
($\sim 10^8M_\odot$; \S~\ref{sec:MCG}) and the galactic black hole
Cyg X-1 ($\sim 30M_\odot$; present section), for the initial 
conditions used
based on properties of the accretion disk,
the main differences in the emitted spectra 
are the number of Penrose produced
$e^-e^+$ pairs escaping, and the range of $E_\mp$: for the 
Seyfert galaxy $E_\mp$ is  in the narrow range $\sim 2.2-2.6$~MeV, and
for the galactic black hole, $E_\mp\sim 0.8-4$~MeV.  In both cases,
most (if not all) of the PPP electrons have turning points in 
 the nonequatorially confined
(spherical-like)  electron orbits at $\sim r_{\rm mb}$,
as discussed in \S~\ref{sec:asymmetry}, indicating that these
electrons escape along vortical trajectories collimated about the
polar axis, without interacting appreciably with the inner edge 
of the bound stable  accretion disk (located at $\sim r_{\rm ms}$;
Williams 2004).

\section{Conclusions}
\label{sec:conclusions}

From the Penrose-Williams model presented here, to extract 
energy momentum
from a rotating black hole, we can conclude the following:
PCS is an effective way to boost
soft X-rays to hard X-rays and  $\gamma$-rays up to $\sim 7- 14$~MeV.
PPP (\gggg) is an effective way to produce
relativistic $e^-e^+$ pairs up to $\sim 10- 54$~GeV:
This is the probable mechanism  producing the fluxes of
relativistic pairs emerging from cores of AGNs; and when
relativistic beaming is included, apparent energies $\sim$~TeV
can be achieved (Williams 2003).
These Penrose processes can operate for any size
rotating black hole, from quasars to
microquasars (i.e., galactic black holes).
Overall, the main features of quasars:
(a) high energy particles (X-rays, $e^-e^+$ pairs, $\gamma$-rays)
coming from the central source;
(b) large luminosities;
(c) collimated jets;
(d) one-sided (or uneven) polar jets---which under certain conditions 
the asymmetry brightness appears to ``flip,''
can all be explained by these Penrose processes. 

Moreover, it is shown here that the geodesic treatment of 
individual particle processes
close to the event horizon, 
as governed by the black hole,
is sufficient to described the motion of the particles.
This is consistent with MHD that  
the behavior of such  individual particles on geometry 
(or gravity)-induced
trajectories is also that of the
bulk of fluid elements in the guiding center approximation
(de Felice \& Zanotti 2000).
In light of this, with some ease, MHD  should be incorporated
into these calculations, particularly to describe the flow of the
Penrose escaping particles away from the black hole,
to perhaps further
collimate and accelerate these jet particles out to the
observed  distances.

Importantly, we can conclude that, the difference between
quasars, radio quiet and radio loud galaxies, and
microquasars, appears to be the 
presence or the lack of a two-temperature ADAF: with or without
nuclear reactions
($pp\ra \pi^0\ra\gamma\gamma$) 
in the inner region of the 
accretion disk (see Eilek 1980;  Eilek \& Kafatos 1983).
Quasars appear to have thin disk/ion corona (ADAF) with 
nuclear reactions. 
In the case of the radio quiet and radio loud galaxies
the  ADAF may no longer be ``nuclear reactive,'' 
however just hot, and in some
cases the disk may have evolved back to its cool 
thin disk phase, including
the associated thermal-cycle Lightman instabilities 
(Lightman 1974a, 1974b).  
The microquasars, on the other hand,
appear in general not to satisfy conditions for the existence of
an ADAF, which is 
determined by the accretion rate (Williams \& Hjellming 2002), but do 
appear to satisfy conditions to have a soft X-ray inner region
and an apparent thermal-cycle instability, with disk 
temperature up to $kT_e\sim 50$~keV.

Finally, what makes the Penrose mechanism described here so 
admirable is that it
allows one to relate the macroscopic conditions, i.e., of the
global gravitational field of the KBH, to the microscopic world of
particle physics.  This description, which is progressively 
being proven by
observations, to be the correct description, allows us to see
directly how
energy is extracted from a black hole.  The physics used in this
Penrose analysis is
that of special and general relativity.
From this analysis and its
consistency with observations,
we arrive at the following conclusion:
Close to the event horizon,  gravity and particle-particle
interactions, in the ergosphere, of highly curved spacetime
(where the effect of the external accretion disk magnetic field
is apparently negligible), are sufficient to described
energy-momentum extraction from a rotating black hole.                               

\acknowledgments

I first thank God for  His thoughts 
 and for making this research possible.  
Next, I thank 
Dr.~Fernando de Felice and Dr.~Henry Kandrup for their helpful comments
and discussions.  Also, I thank Dr.~Roger Penrose for his continual
encouragement.  I am grateful to the late Dr.~Robert (Bob) Hjellming for
his helpful discussions and cherished collaboration.  Part of this 
work was done at the Aspen Center for Physics.  This work was supported 
in part by a grant from NSF at NRAO and an AAS Small Research Grant.

\appendix
\section*{APPENDIX}
\label{sec:appendix}

Associated problems with popular MHD models are described
below:
 
1. In order to explain observations of the
Seyfert 1 galaxy MCG---6-30-15, that copious
photons are been  extracted from the black hole from radii
less than the marginal stable orbit $r_{\rm ms}$
($\simeq 1.2M$, in gravitational
units with $G=c=1$, where $M$ is the mass of the black hole),
it has been
claimed that the force lines of the disk magnetic
field ${\bf B}_d$ couple with
matter deep within
the ``plunging'' region $< r_{\rm ms}$, thereby extracting
rotational energy in the form of electromagnetic energy
(Wilms et al. 2001; Krolik 2000).  However, the first detailed
numerical relativistic
time-dependent MHD  calculations
in a Kerr metric (Koide et al. 2000; Meier \& Koide 2000;
Meier, Koide, \& Uchida 2001) show that
in order for magnetic
field lines to
extend inward to the numerical limited radius
$ 1.3 M$---being
frozen to the plasma,
of Keplerian velocity, the disk material must be initially
counter rotating:  opposite the direction that the black
hole is rotating.
This appears inconsistent with the
observations of Zhang, Cui, \& Chen (1997)
and in general the physics
occurring inside the ergosphere in which inertial frames
are dragged in the direction that the black hole is rotating.
Even though we know that particles
can have retrograde orbits inside the ergosphere, relative to
an observer at infinity,
it is highly improbable that the whole
disk of matter will be counter rotating, at least in the general sense. 
Further, it appears that the net rotational
energy being ``extracted'' in the numerical simulation of these
authors (Koide et al. 2000) in the form of electromagnetic energy
over and above the gravitational binding energy released due to
 the hydrodynamic energy transported into the
black hole is merely
the rotational energy from the nonphysical initial condition that the
accretion disk plasma is counter rotating as it falls into the
ergosphere.\footnote{Moreover, these authors (Koide et al. 2000) made the
statement that inside the ``static limit''
(i.e., ergosphere), the velocity of the frame dragging exceeds the speed
of light ($c\Omega_3/\alpha>c$)! Not only is this an untrue statement,
but it is a violation of the laws of physics.   The frame dragging
circular velocity inside the ergosphere as measured by an observer
at infinity is $\omega\sqrt{g_{\Phi\Phi}}
\sim 0.8-0.9c$ (see Bardeen et al. 1972; Misner et al. 1973;
Thorne et al. 1986; Williams 1995; see also \S~\ref{sec:cyg}).}
On the other hand, for a co-rotating disk these authors found that the
inward limiting radius is even larger ($\sim 6M$), attributed to a
 centrifugal barrier (Koide et al. 2000).
Although this time-dependent MHD model is an excellent representation
of  subrelativistic (${<\atop\sim\,} 0.4c$) jet formation in a KBH
magnetosphere, the inconsistencies of this MHD model, as matter nears
the event horizon ($r_+\simeq 1.063 M$), is probably an indication of
the limitation, of such fluid dynamical models, in describing energy
extraction from a rotating black hole: this being  based on the
guiding center
approximation,  wherein
the single-particle approach is essential
close to the black hole (de Felice \&  Carlotto 1997; de Felice \&
Zanotti 2000), i.e., the behavior of individual particles
is also that of the bulk of fluid elements.   This means that
gravitational-particle interactions,
such as the Penrose processes
describe here (in this paper), are required.
Note, in these Penrose processes, which occur close to the event 
horizon, elementary electromagnetic and atomic
forces dominate on the microscopic scale, while gravity is dominant
on the macroscopic scale---thus, as it should be in the strong
gravitational potential well of the KBH; but far away
from $r_+$ electromagnetism
appears to dominate macroscopically (Junor, Biretta, \& Livio 1999).
Moreover, stability of the
co-rotating disk, falling inward to the limiting
radius $\sim 6M$,  at the Keplerian velocity,
when  magnetic field lines are coupled to the
infalling plasma, with the jet formation similar to that of the
Schwarzchild black hole case (Koide, Shibata, \& Kudoh 1999),
suggests that the large scale magnetic field plays a dominant
role at large distances from $r_+$, irrespective of whether or
not the black hole is rotating.
In addition, these numerical inward
limiting radii, at least
in the case of the counter-rotating disk ($\sim r_{\rm ms}$),
may also be a display of the
horizon being a ``vacuum infinity'' (Punsly \& Coroniti  1989;
Punsly 1991; Williams 2003): to the associated magnetic field
charge neutral disk particle plasma, in accordance with the 
``no-hair'' theorem
(Carter 1973; Misner, Thorne, \& Wheeler 1973; Williams 1995),
suggesting that the interaction of the disk magnetic  field with
particles in bound, trapped  orbits at radii $< r_{\rm ms}$ is
negligible compared to the Penrose gravitational-particle
interactions described here.
Therefore, it appears that electromagnetic
energy cannot be effectively
extracted from the so-called plunging region: where
gravitational-particle interactions will clearly dominate
if the magnetic flux of an axisymmetric ${\bf B}_d\ra 0$, 
as it does in general upon nearing the  vacuum
infinity horizon [Punsly \& Coroniti  1989;
Punsly 1991; Williams 2003; see also Bi\v{c}\'{a}k (2000) and
Bi\v{c}\'{a}k \& Ledvinka (2000) for a detailed general relativistic 
calculation showing this].

2. To convert the electromagnetic
energy to
particle energy at the event horizon, and to
duplicate the observed luminosities from a Poynting flux,
it requires a large-scale magnetic field strength
$B_d\sim 10^4 (M/10^7M_\odot)^{-1}$~G
(Wilms et al. 2001; Blandford \& Znajek 1977). In order to
create $e^-e^+$ pairs
along the field lines, as in the case of pulsars, a field
strength of at least $B_d\sim 10^{12}$~G is needed (Sturrock 1971;
Sturrock, Petrosian, \& Turk 1975).
(The mechanism, however, for the generation of the pairs in an
electromagnetic field to date is a subject of debate.)
The first of the large strengths required above
appears to be achieved for
supermassive KBHs, at present---i.e., with speculated assumptions. But 
for  galactic
black holes  (microquasars) with masses
$\sim 10M_\odot$,
$B_d\sim 10^{10}$~G seems highly impossible to generate from,
in most cases, a binary system accretion
disk plasma flow.  An effective
model for AGNs must also operate for
microquasars as well.  Moreover,  according to
electrodynamics, in general, to
 lift the particles ``frozen'' to the magnetic
field lines,
from a disk, accelerating them to relativistic speeds, there has
to be an electric field component
$ E_z$ (Lovelace 1976).
However, there
exist  problems in generating sufficient $\bf E$ parallel
to the polar direction
($\pm\,{\bf  e_z}$ axis); none of the polar MHD models
of this particular type adequately gets rid of this problem.  
Magnetic reconnection may be
a solution to some degree.

3. To get around problems in items 1 and 2 (specifically, the large
strength field required and the vacuum infinity horizon)
it is assumed that a ``hot'' ion corona or torus-like accretion
can provide
the necessary jet particles: ($a$) for the magnetosphere to act on,
accelerating and collimating through centrifugal driving winds (see below);
and ($b$) to provide the hot ram pressure, to
``ram'' the magnetic field lines inward to the event horizon.
However, now there appears to be a problem
concerning how to
liberate particles from trapped orbits inside the ergosphere
(particularly in the plunging regime) onto escaping orbits.
Particles in the plunging regime,
as defined by Bardeen, Press, and Teukolsky (1972), i.e.,
massless and material particles (with $E/\mu_o\geq 1$) originating from 
infinity,
can only escape, by being injected onto
escaping orbits by some physically process near the black
hole---such as the Penrose scattering processes described
here---since nothing can come out of the
hole (Bardeen et al. 1972).
Therefore,  the BZ-type models are
 faced with yet another problem, as the
magnetic field is assumed to get closer to the KBH: where
general relativistic
effects must be considered, i.e., how do we get the necessary
escaping particles in numbers out of the ergospheric region
($< r_{\rm ms}$)
 into the jets by such models?
Moreover, with observations showing M87
not having the expected large ``dusty'' thermal IR-emitting torus
(Perlman et al. 2001) that could have possibly served as
particle jet ``fuel'' for a BZ-type model,
the Penrose
mechanism to extract energy momentum, as described by Williams
(1995), the so-called Penrose-Williams mechanism,
appears to be
the only  possible,
plausible way to power this AGN, and thus, generate its
jets (\S~\ref{sec:M87}).
So, in summary, in addition to the problems above associated
with the BZ-type
MHD models, there still exists the historical problem: How does one
convert from
electromagnetic energy  to the 
particle energies observed in the jets, emanating from the region
where energy is observed to be
extracted, i.e., inside the ergosphere close to the event horizon?
None of the existing BZ-type MHD models thus far adequately solves
this ``age-old'' problem.
 
4. In the centrifugal driven winds (Blandford \& Payne 1982)
mentioned above, the following is assumed: If the disk magnetic
field lines subtends an angle of more than $\pm 30^\circ$ to the
rotation axis, the gas will be flung away from the disk into
collimated jets with speeds a few times the escape velocity at the
magnetic footprint on the disk.  Now, this may be true at $r\gg r_+$, but
near the event horizon $r_+$, the escape conditions (see Williams
1995) must be adequately applied. Recently, a  general relativistic MHD
treatment of evolving tori that includes in some degree features 
of the Blandford \& Payne (1982) type-models, which  
allow for centrifugal driven winds to power
the jets (Hirose et al. 2004; De Villiers, Hawley, \& Krolik
2004; De Villiers, Hawley, \& Krolik 2003), found no such appreciable 
relativistic winds emerging
from the horizon, nor the so-called plunging region, nor
the ergospheric accretion disk that could
be tied directly to rotational energy extraction from the black hole,
although it was found that the Lorentz force
inside the ergosphere increased due to inertial frame dragging.   
The plunging region for $a/M=0.998$ lacked adequate resolution, 
suggesting perhaps the need for a general relativistic particle geodesic
treatment, according to the guiding center approximation
(discussed in item~1 of this Appendix; see also Williams 2003).  
These MHD calculations (Hirose et al. 2004; De Villiers et al.
2004; De Villiers et al. 2003) did, however, confirm the existence of 
the predicted funnel region (Rees et~al. 1982; see Williams 2003), and  
are consistent with the evolved magnetic field configuration
found by Bi\v{c}\'{a}k  (2000) and  Bi\v{c}\'{a}k \& Ledvinka
(2000), i.e., that radial lines are expelled from the surrounding 
equatorial region (${\bf B}_r\rightarrow 0$), but at the poles
${\bf B}_r\neq 0$.  This clearly shows that the classical BZ-type
models (Blandford \& Znajek 1997), where magnetic field lines are 
proposed to anchor to the 
event horizon, thereby extracting rotational energy, could not be
an important source, because ${\bf B}_r\rightarrow 0$ in the region
of importance for extracting rotational energy.  Also these MHD
calculations
seem to confirm the importance of magnetic fields on a 
large scale,
in gravitational accretion processes, i.e., in aiding 
mass outflows to large 
distances in the jets of black holes as well as those of 
protostars (see Williams 2004).

5. Finally, to clear up any confusion,
the authors of the historical paper (Wilms et al. 2001) loosely
called the
BZ-type models
the Penrose effect---the very name for years that had distinguished
Williams' (1991, 1995, 1999, 2001) internationally known
successful four-dimensional
Penrose  model (see also Piran \& Shaham 1977; Leiter \& Kafatos
1978; Kafatos \& Leiter 1979; Kafatos 1980; Wagh \& Dadhich 1989) from
the BZ-type models.
Strangely, these authors did not
reference Williams'  investigation.
Nevertheless, to set the record straight,
the Penrose mechanism
[as summarized here and described in detail in Williams (1995)],
which involves
gravitational extraction of energy from a spinning black hole, 
based on that visualized by Penrose (1969),
and that of
the so-called BZ mechanism, which involves
electromagnetic extraction of energy (Blandford \& Znajek 1977),
are two very different models. So different that the statement
made by the authors in Wilms et al. (2001), ``For parameters
relevant to our discussion, the extra energy source is provided by the
spin via the Penrose effect occurring within the radius of
marginal stability (but outside of the stretched horizon),'' indeed
requires a proper reference, since Williams' (1991, 1995)
model is popularly known
as the only existing completely worked out model of the
Penrose mechanism: occurring within the radius of
marginal stability $r_{\rm ms}$.
Whatever the case may be, the
recent observations of MCG---6-30-15 (Wilms et al. 2001) and M87
(Perlman et al. 2001)
introduce compelling evidence suggesting
that perhaps it is the effects of Williams'  
black hole source
model that is being observed (as 
described in this paper), and hardly those of the BZ-type models.
The evidence presented here strongly suggests that observed black 
hole sources have a central energy generation similar the mechanism 
described in this present paper.  So, to avoid any further 
confusion, it seems appropriate to refer to Williams' model as the
Penrose-Williams mechanism, which I interchangeably
refer to as just the Penrose mechanism, out of respect for its
originator Penrose (1969).

%References:

\clearpage

\begin{table}
\begin{center}
\caption{Model Parameters for 3C 273 (PCS)\label{tbl-1}}
\begin{tabular}{crrrrrrrr}
\tableline\tableline
 &$r$~~~&$E_e$~~~~&$~~\log (\nu_{\rm ph})$&
$\log (\nu_{\rm peak})$
&$\log (L_{\rm peak})$&$\log (L_{\rm obs})$
&$f_1~~~~$
&$f_2$~~~~~~~ \\
Case no. &$(M)$&(MeV)&(Hz)
&(Hz)&
$({\rm erg/ s})$&$({\rm erg/ s})$&&\\
\tableline
$1^{\rm \,a}$...... &$1.089  $&$~0.539^{\rm \,b} $&$16.23 $&$18.24 $
&$~~45.93$&$\ldots$~~~&$4.09\, (-2) $&$1.0\,$~~~~~~~~~~~~\\
2........&$1.099 $&$0.456~~ $&$17.09  $&$18.91 $
&$~~46.20$&$46.0~ $&$4.17\, (-2) $&$~~~1.0$ ~~[0.632]\\
5........&$1.089  $&$0.539^{\rm \,b} $&$17.86 $&$19.72 $
&$~~46.61$&$46.4~$
   &$4.09\, (-2) $&$~~~1.0$ ~~[0.611]\\
6........&$1.089  $&$0.539^{\rm \,b} $&$18.86 $&$20.21 $
&$~~46.27$&$\ldots$~~~&$4.09\, (-2) $&$~1.0\,$~~~~~~~~~~~~\\
7........&$1.089  $&$~0.539^{\rm \,b} $&$19.56 $&$20.56 $
&$~~45.48$&$\ldots$~~~
   &$4.09\, (-2) $&$~1.0\,$~~~~~~~~~~~~\\
$8^{\rm \,c}$......&$1.089  $&$1.435~ $&$18.86 $&$20.61 $
&$~~47.12$&$46.2~$
   &$4.09\, (-2) $&$~~~1.0$ ~~[0.121]\\
11........&$1.089  $&$4.543~ $&$18.86 $&$21.06 $
&$~~48.37$&$46.08$
   &$4.09\, (-2) $&$~~~1.0$ ~~[0.005]\\
13........&$1.089  $&$11.79\,~~ $&$18.86 $&$21.46 $
&$~~49.34$&$~~46.08
  $ &$4.09\, (-2) $&$~~~1.0$ ~~[0.001]\\
\tableline
\end{tabular}
\tablenotetext{a}{Case numbers 1 through 7 are for PCS by
equatorially confined target
electrons}
\tablenotetext{b}{When the more exact value is used for $r=r_{\rm mb}
=1.091M$, $E_e\lra 0.512$~MeV $\simeq \mu_e$ (see Williams 1995,
2003; Bardeen et al. 1972),
as would be expected for equatorially
confined orbits}
\tablenotetext{c}{Case numbers 8 through 13 are for PCS by nonequatorially
confined target electrons}
\end{center}
\end{table}
\clearpage

\begin{table}
\begin{center}
\caption{Model Parameters for 3C 273 (PPP)\label{tbl-2}}
\begin{tabular}{crrrrrrrr}
\tableline\tableline
 $r=r_{\rm ph}$ &$(E_{\mp})_{\rm peak}$&$\log (\nu_{\rm peak})$
&$\log (L_{\rm peak})$
&$\log (L_{\rm obs})$
&$f_1=f_3$
&$f_2=f_4~~$&$f_5~~~~~~$\\ 
Case no. &~(MeV)&~(Hz)
&$~~({\rm erg/ s})$&
$~~({\rm erg/ s})$&&&\\
\tableline
14$^{\rm a}$........ &$~~~1.289$ &$20.63 $&$45.18 $
&$\ldots$~~~&$~1.99\, (-2)$
&$1.0\,~~~~~~~~~~~$&$1.0\,~~~~~~~~~~~$\\
15........&$~~~6.626 $&$21.21 $&$46.32 $
&$~~46.06~$
&$~1.99\, (-2) $&$1.0$ ~~[0.8]~~~~&$~~1.0$ ~~[0.859]\\
17........&$73.37~~ $&$22.25 $&$48.58 $
&$45.7~~\,  $
&$~1.99\, (-2) $&$~~1.0$ ~~[0.1]~~~~&$~~1.0$ ~~[0.132]\\
19........&$174.6~~~ $&$22.59 $&$49.36 $
&$45.6 ~~\,$
&$~1.99\, (-2) $&$~~~1.0$ ~~[0.05]\,~~&$~1.0$ ~~[0.069]\\
22........&$711.8~~~$&$23.21 $&$50.65 $
&$~~45.25~$
&$~1.99\, (-2) $&$~~~~1.0$ ~~[0.022]&$~1.0 $ ~~[0.008]\\
25........&$2469\,~~~~~$&$23.77 $&$51.85$
&$44.8~~\, $
&$~1.99\, (-2) $&$~~~~1.0$ ~~[0.008]&$~1.0$ ~~[0.001]\\
\tableline
\end{tabular}
\tablenotetext{a}{ Case numbers 14 through 25 have infalling initial
(incident)  photon frequency, used in the ``secondary Penrose
Compton scattering'' (SPCS),  $\nu_{\rm ph}\simeq 7.24\times
10^{18}$~Hz}
\end{center}
\end{table}                                                                            
\clearpage

%Figure Captions and Figures:
%\begin{figure}[t!] % fig 1
\begin{figure} % fig 1
\epsscale{.65}
\plotone{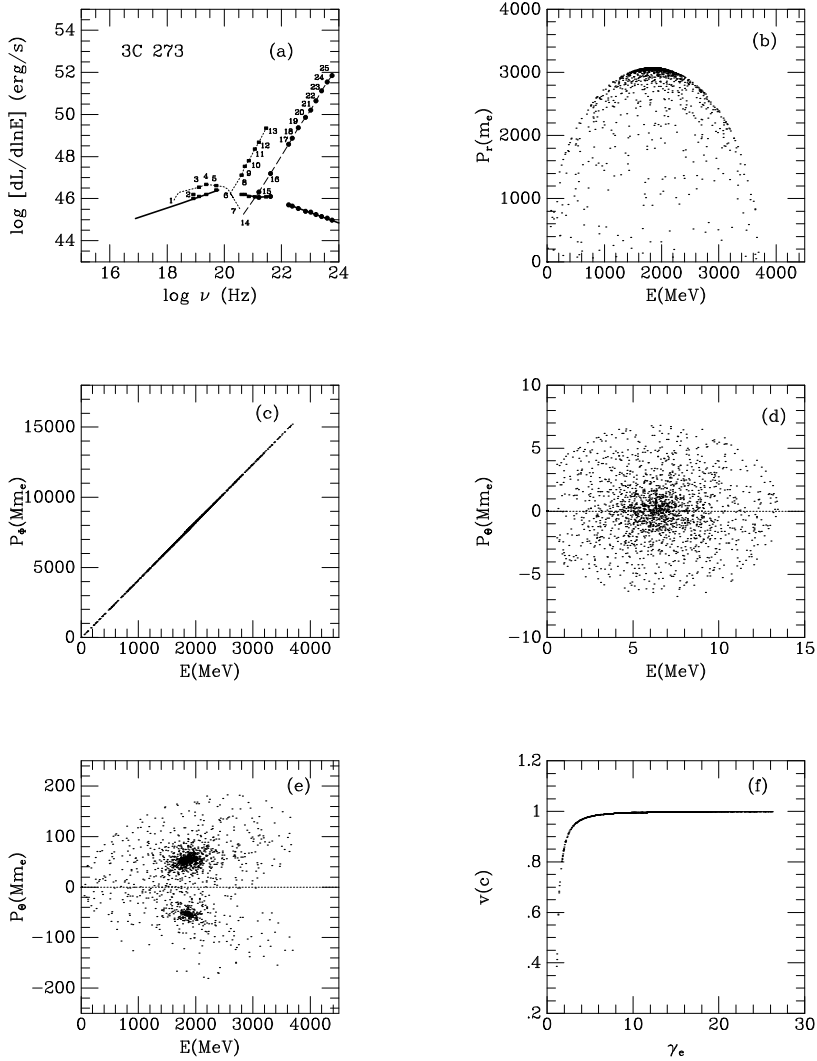}
%\vspace{35pt}
%\vskip 3.0in
\caption{(a) Comparing the theoretical spectrum
with observations for 3C~273. The calculated
PCS and PPP (\gggg) luminosity spectra are represented by the solid
squares and large solid dots, respectively.
The observed spectra is indicated
by the solid line.  The upper curves with the solid squares and
solid dots superimposed on the dotted line and the
dashed line, respectively,
for PCS and PPP (\gggg), are the general 
spectra calculated from this model.
Superimposed on the lower solid line of the observations are
solid squares and solid dots that have been fitted to agree with
observations.   These fits depend on the $f_n$'s values (see
text). (b) and (c) PPP
(\gggg) at $r_{\rm ph}=1.074M$:
scatter plots showing
momentum components of the escaping $e^-e^+$ pairs
(each point represents a particle):
radial momenta $(P_\mp)_r$
vs.~$E_\mp$ and azimuthal coordinate
momenta $(P_\mp)_\Phi~(\equiv L_\mp)$ vs.~$E_\mp$,
respectively;
for the infalling photons
$E_{\gamma 1}=0.03$~MeV, and for the target photons
$E_{\gamma 2}\simeq 3.893$~GeV, $(P_{\gamma 2})_\Th=\pm 113\,Mm_e$.
(d) and (e) PPP (\gggg):  polar coordinate momenta 
$(P_\mp)_\Th $ vs.~$E_\mp $
for  $E_{\gamma 1}=0.03$~MeV, $E_{\gamma 2}\simeq 13.54$~MeV,
$(P_{\gamma 2})_\Th=\pm 0.393\,Mm_e$
 and for $E_{\gamma 1}=0.03$~MeV,
$E_{\gamma 2}\simeq 3.893$~GeV, $(P_{\gamma 2})_\Th=\pm 113\,Mm_e$,
respectively. (f) The velocity distribution vs.
$\gamma_e$($={E_\mp/ m_ec^2}$) for the same case as (d) above.
Note, $M=10^8M_\odot$.
 \label{fig1}}
\end{figure}
\clearpage

\begin{figure} % fig 2
\epsscale{.4}
\plotone{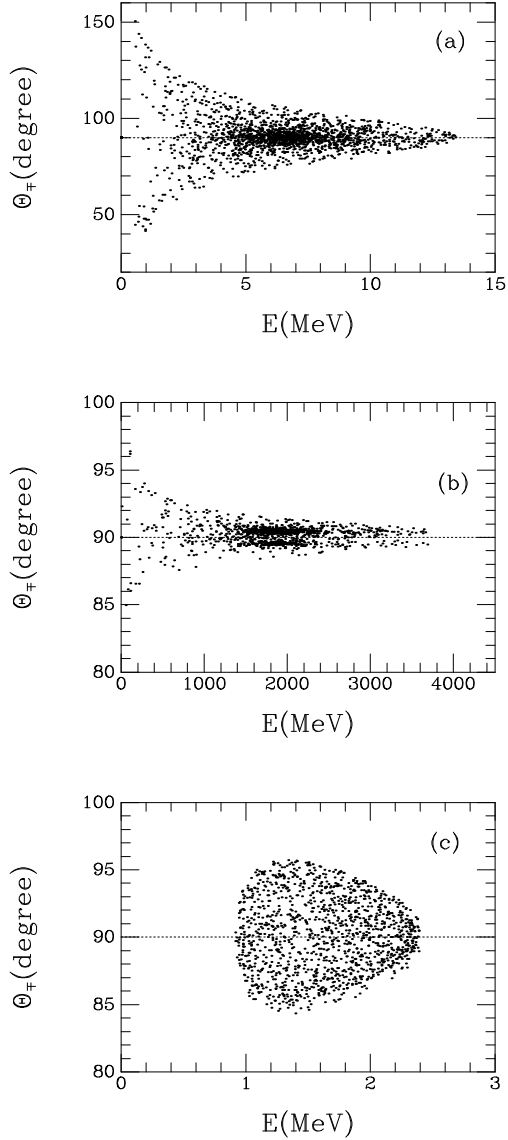}
%\vspace{15pt}
%\vskip 2.5in
\caption{PPP
(\gggg) at $r_{\rm ph}=1.074M$, for $M=10^8 M_\odot$:
scatter plots displaying polar angles, above and below 
the equatorial plane:
$\Theta_\mp$ vs.~$E_\mp$, of the escaping
$e^-e^+$ pairs after 2000 events (each point represents a 
particle).
The cases shown are defined
by the following
parameters: $E_{\gamma 1}$, the infalling photon energy;
$E_{\gamma 2}$, the target photon orbital energy;
$Q_{\gamma 2}^{1/2}$, corresponding polar coordinate
momentum $(P_{\gamma 2})_\Theta$ of the target photon;
$N_{\rm es}$, number
of  $e^-e^+$ pairs escaping.
(a) $E_{\gamma 1}=0.03$~MeV,
 $E_{\gamma 2}\simeq 13.54$~MeV,
$Q_{\gamma 2}^{1/2}=\pm 0.393\,Mm_e$,
$N_{\rm es}=1850$.
(b) $E_{\gamma 1}=0.03$~MeV,  $E_{\gamma 2}\simeq 3.893$~GeV,
$Q_{\gamma 2}^{1/2}=\mp 113.0\,Mm_e$,
$N_{\rm es}=1997$.  (c) $E_{\gamma 1}=3.5$~keV, 
$E_{\gamma 2}\simeq 3.4$~MeV,
$Q_{\gamma 2}^{1/2}=\mp 0.0987\,Mm_e$,
$N_{\rm es}=1326$.  Note that, $\Theta_\mp > \pi/2$
is below the equatorial plane. \label{fig2}}                        
\end{figure}
\clearpage

\begin{figure} % fig 3
\epsscale{.7}
\plotone{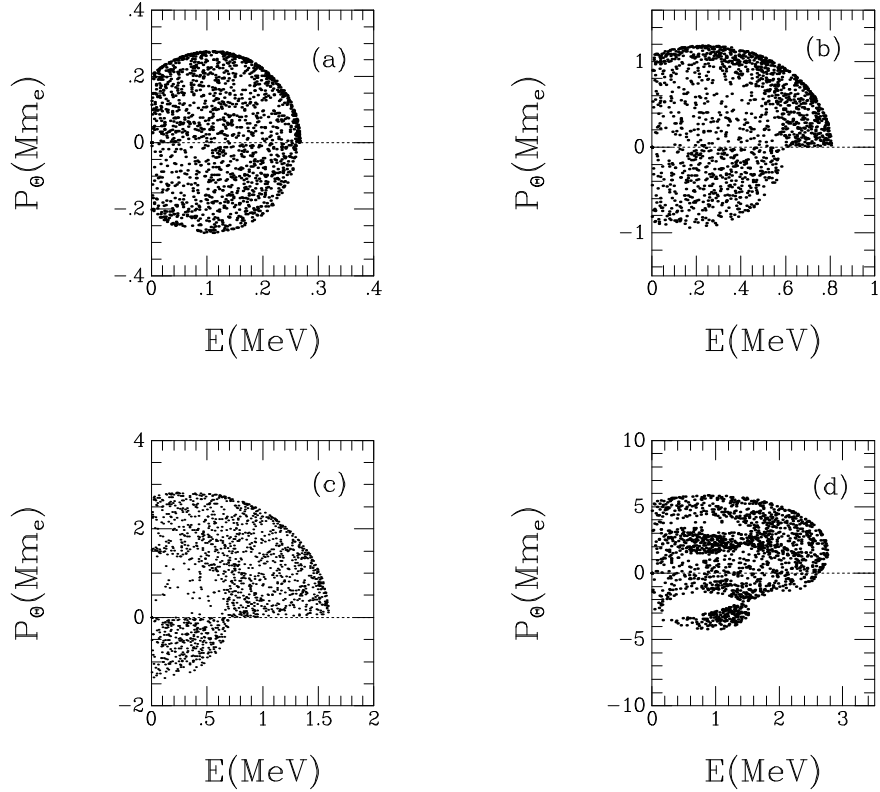}
%\vspace{10pt}
%\vskip 0.5in
\caption{PCS: scatter plots showing
polar coordinate space momenta: $(P_{\rm ph}^\prime)_\Theta$
$[\equiv (Q_{\rm ph}^\prime)^{1/2}$]
vs.~$E_{\rm ph}^\prime$, of the escaping PCS photons after 2000 events
(each point represents a
particle), at $r_{\rm mb}\simeq 1.089M$, for $M=10^8M_\odot$.
The various cases are defined
by the following
parameters: $E_{\rm ph}$, initial photon energy; 
$E_e$, the target electron orbital energy;  $Q_{\rm e}^{1/2}$, 
defining the corresponding polar coordinate
momentum $(P_{e})_\Theta$ of the target electron;
$N_{\rm es}$, number
of photons escaping. (a) $E_{\rm ph}=3.5$~keV,
 $E_e\simeq 0.539$~MeV, $Q_{\rm e}^{1/2}=0$,
$N_{\rm es}=1637$.
(b) $E_{\rm ph}=0.03$~MeV, 
 $E_e\simeq 0.539$~MeV, $Q_{\rm e}^{1/2}=0$,
$N_{\rm es}=1521$.
(c) $E_{\rm ph}=0.15$~MeV, $E_e\simeq 0.539$~MeV, 
 $Q_{\rm e}^{1/2}=0$,
$N_{\rm es}=1442$.
(d) $E_{\rm ph}=0.15$~MeV,
 $E_e\simeq 1.297$~MeV, $Q_{\rm e}^{1/2}=\pm 2.479\,Mm_e$,
$N_{\rm es}=1628$. [Note, due to a minor
 oversight leading to improper treatment in the computer simulation
of the arccosine term in
 eq.~(3.39) of Williams (1995),
 correct Figs.~3$a$ and~3$b$ presented here
replace Figs.~7(a)
and~3(c), respectively, of Williams (1995).] \label{fig3}}
\end{figure}
\clearpage

\begin{figure} % fig 4
\epsscale{.7}
\plotone{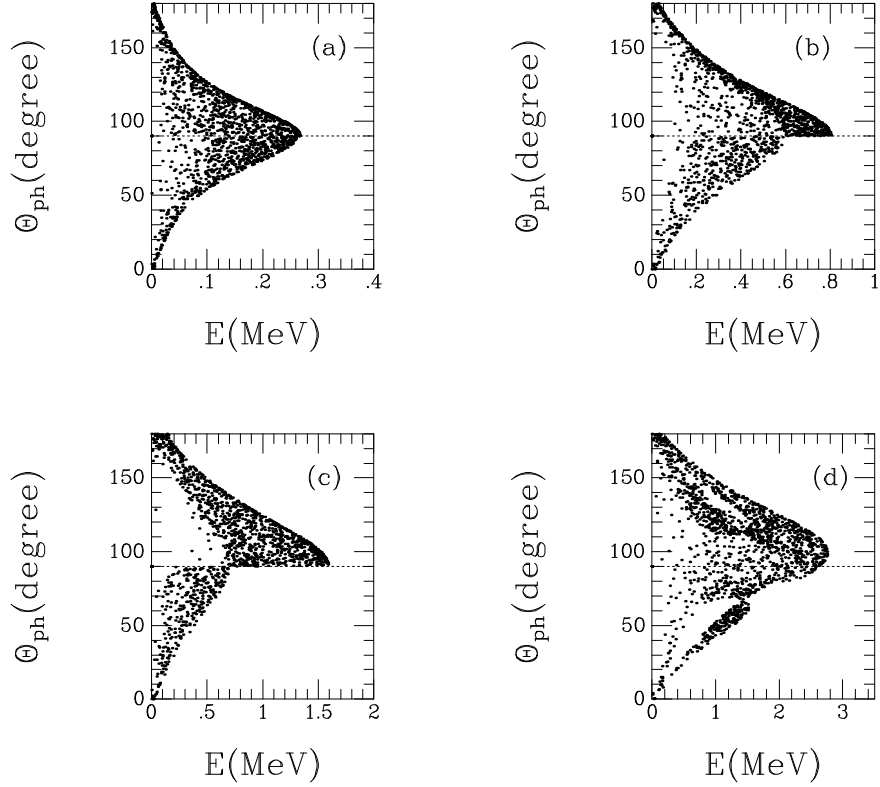}
%\vspace{10pt}
%\vskip 0.4in
\caption{PCS: scatter plots
displaying polar angles, above and below the equatorial plane:
$\Theta_{\rm ph}^\prime$ vs.~$E_{\rm ph}^\prime$, of the escaping
PCS photons,
for the cases $4a-4d$,
 described in Figs.~$3a-3d$, respectively.
Note that, $\Theta_{\rm ph}^\prime > \pi/2$
is below the equatorial plane.  \label{fig4}}
\end{figure}            

\begin{figure} % fig 5
\epsscale{.7}
\plotone{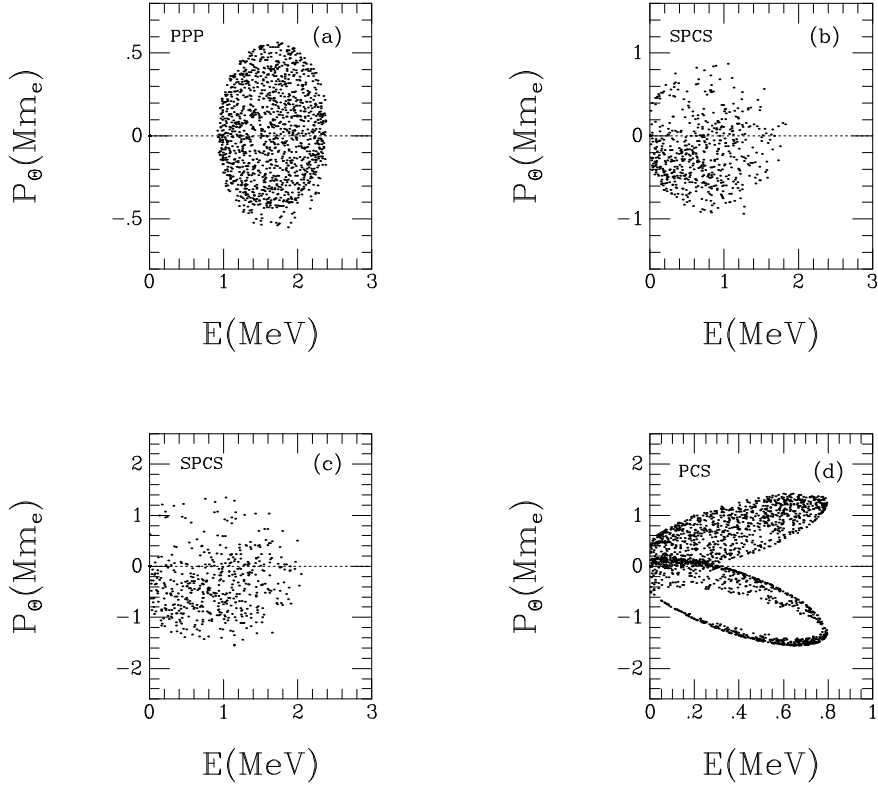}
%\vspace{15pt}
%\vskip .5in
\caption{Self-consistent PPP
(\gggg) and secondary Penrose Compton scattering (SPCS) 
at $r_{\rm ph}=1.074M$; and PCS by nonequatorially confined 
electron targets at $r_{\rm mb}=1.089M$; with $M=30M_\odot$:
scatter plots displaying the polar coordinate momenta, above and 
below the equatorial plane, versus the energy of the 
escaping particles,
per 2000 infalling disk photons for each case shown 
(each point represents a particle from the 
scattering events).
(a) PPP: $(P_\mp)_\Th$ $[\equiv (Q_\mp)^{1/2}$] vs.~$E_\mp$, 
with $E_{\gamma 1}=3.5$~keV,
 $E_{\gamma 2}\simeq 3.4$~MeV,
$Q_{\gamma 2}^{1/2}=\pm 0.125\,Mm_e$;
$\ep_\mp=700/617$ (see text). (b) SPCS: $(P_{\rm ph}^\prime)_\Theta$
$[\equiv (Q_{\rm ph}^\prime)^{1/2}$]
vs.~$E_{\rm ph}^\prime$, with $E_{\rm ph}=3.5$~keV, see Fig.~5$a$ for
range of $E_\mp$ and 
 $Q_\mp^{1/2}\equiv (P_\mp)_\Th$; 
$\ep_{\mp({\rm ph})}=165/402$ (jet reversal; see text).  
(c) SPCS: $(P_{\rm ph}^\prime)_\Theta$
vs.~$E_{\rm ph}^\prime$, with $E_{\rm ph}=10$~keV, see Fig.~5$a$ 
for range of $E_\mp$ and
 $Q_\mp^{1/2}\equiv (P_\mp)_\Th$; $\ep_{\mp({\rm ph})}=127/363$
(jet reversal). 
(d) PCS: $(P_{\rm ph}^\prime)_\Theta$
vs.~$E_{\rm ph}^\prime$, with $E_{\rm ph}=3.5$~keV,
 $E_e\simeq 1.22$~MeV, $Q_{\rm e}^{1/2}=\pm 2.3Mm_e$; 
$\ep_{\rm ph}=1136/706$ (see text). \label{fig5}}
\end{figure}
\clearpage

\begin{figure} % fig 6
\epsscale{.8}
\plotone{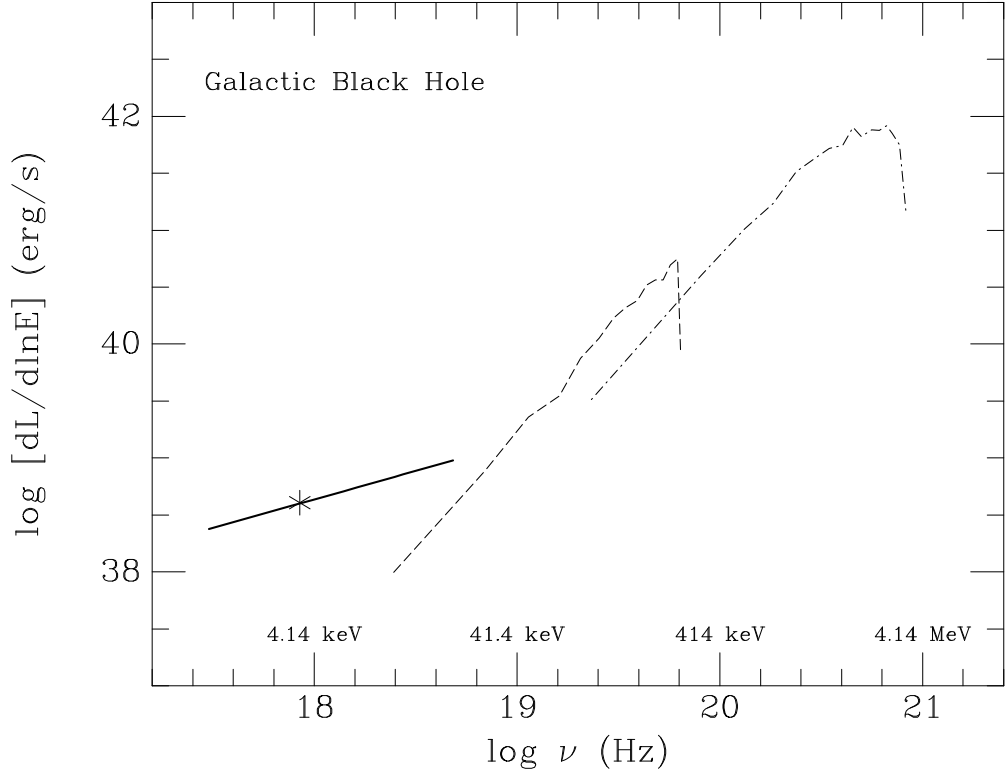}
%\vskip .5in
\caption{Self-consistent luminosity spectra of PCS by 
equatorially confined   
 (dashed curve)
electron targets at $r_{\rm mb}=1.089M$, and
secondary Penrose Compton scattering (SPCS) by PPP electrons
at $r_{\rm ph}=1.074M$ (dashed-dotted curve), with $M=30M_\odot$.   
The total emitted spectrum is similarly to 
that  observed for Cyg X-1 [$M\sim 10M_\odot$ (Liang 1998)].  
The assumed power-law
distribution accretion disk for the inner region
($K\nu^{-\alpha}$, where $\alpha=1.5$; $K=1.1\times 10^9 $~(cgs units), 
in the general range 
($\sim 1.25-20$~keV), is shown (solid curve): the asterisk indicates
monochromatic infalling photon energy producing the self-consistent
Penrose processes displayed [$(L_\gamma)_{\rm in}\sim 4\times 
10^{38}~{\rm erg\,s^{-1}}$; see text].  For PCS by equatorially confined 
targets: $E_{\rm ph}=3.5$~keV,
$E_e\simeq 0.539$~MeV, $Q_e=0$.
For SPCS by PPP electron targets: $E_{\gamma 1}=3.5$~keV, 
$E_{\gamma 2}\simeq 5.012$~MeV, $Q_{\gamma 2}^{1/2}=\pm 0.185Mm_e$.
 \label{fig6}}
\end{figure}
                                          
\end{document}